\documentclass[manuscript]{aastex}
\usepackage{graphicx}
\usepackage{dcolumn}
\usepackage{bm}

\newcommand{\ignore}[1]{}

\newcommand{\grad}{\mbox{grad}}

\newcommand{\pdiffl}[2]{\frac{\partial #1}{\partial #2}}
\newcommand{\diffl}[2]{\frac{d #1}{d #2}}

\newcommand{\SIdens}{g~cm$^{-3}$}

\newcommand{\ME}{M$_{\mbox{E}}$}
\newcommand{\RE}{R$_{\mbox{E}}$}
\newcommand{\MJ}{M$_{\mbox{J}}$}
\newcommand{\RJ}{R$_{\mbox{J}}$}

\begin{document}


\title{Mass-radius relationships for exoplanets}

\date{January 26, 2010, modified April 13, 2011 and September 12, 2011
-- LLNL-JRNL-421766}

\author{D.C.~Swift, J.H.~Eggert, D.G.~Hicks, S.~Hamel, K.~Caspersen,
   E.~Schwegler, G.W.~Collins}
\affil{%
   Lawrence Livermore National Laboratory,
   7000 East Avenue, Livermore, California 94550, U.S.A.
}
\author{N.~Nettelmann}
\affil{%
   Institut f{\"u}r Physik, Universit{\"a}t Rostock, D-18051 Rostock, Germany
}
\author{G.J.~Ackland}
\affil{%
   Centre for Science at Extreme Conditions,
   School of Physics,
   University of Edinburgh,
   Edinburgh, EH9~3JZ, Scotland, U.K.
}

\begin{abstract}
For planets other than Earth, particularly exoplanets,
interpretation of the composition and structure depends largely on
comparing the mass and radius with the composition expected given their
distance from the parent star.
The composition implies a mass-radius relation which relies heavily on equations of state calculated
from electronic structure theory and measured experimentally on Earth.
We lay out a method for deriving and testing equations of state, 
and deduce mass-radius and mass-pressure relations for key, relevant materials 
whose equation of state is reasonably well established,
and for differentiated Fe/rock.
We find that variations in the equation of state,
such as may arise when extrapolating from low pressure data,
can have significant effects on predicted mass-radius relations,
and on planetary pressure profiles.
The relations are compared with the observed masses and radii of
planets and exoplanets, broadly supporting recent inferences about
exoplanet structures.
Kepler-10b is apparently `Earth-like,' likely with a proportionately larger
core than Earth's, nominally 2/3 of the mass of the planet.
CoRoT-7b is consistent with a rocky mantle over an
Fe-based core which is likely to be proportionately smaller than Earth's.
GJ~1214b lies between the mass-radius curves for H$_2$O and CH$_4$,
suggesting an `icy' composition with a relatively large core or a
relatively large proportion of H$_2$O.
CoRoT-2b is less dense than the hydrogen relation, which could be explained
by an anomalously high degree of heating or by higher than assumed atmospheric opacity.
HAT-P-2b is slightly denser than the mass-radius relation for hydrogen,
suggesting the presence of a significant amount of matter of higher atomic
number.
CoRoT-3b lies close to the hydrogen relation.
The pressure at the center of Kepler-10b is $1.5_{-1.0}^{+1.2}$\,TPa.
The central pressure in CoRoT-7b is probably close to 0.8\,TPa,
though may be up to 2\,TPa.
These pressures are accessible by planar shock and ramp loading 
experiments at large laser facilities.
The center of HAT-P-2b is probably around 210\,TPa,
in the range of planned 
National Ignition Facility experiments, and that of CoRoT-3b around 1900\,TPa.
\end{abstract}


\maketitle

\section{Introduction}
Planets outside the solar system have been detected since 1992
\citep{Wolszcan1992} from the Doppler shift of spectral features in emission from
the star, which determines the orbital period and places a constraint
on the mass of the planet \citep{Mayor1995}.
Since 1999, the presence of exoplanets has also been deduced from their transit
across the face of the parent star \citep{Henry2000}.
The fraction of light blocked by the planet allows the radius of the planet
to be deduced as a function of the radius of the star 
\citep{Charbonneau2000,Carter2011}.
Subsequently, several hundred exoplanets have been detected
at different distances from their stars,
and the precision with which mass and radius have been deduced has increased
for some exoplanets to better than 10\%\ in both mass and radius 
\citep{Schneider2011}.

In 2001, light from an exoplanet was detected directly \citep{Charbonneau2002},
opening the window to studies of exoplanet structure through
the composition and temperature of the surface or atmosphere.
However, inferences about the composition and structure rely on
the comparison of mass and radius with planets within the solar system.
With the exception of Earth, which is currently the only planet
for which seismic data exist, interpretations of the internal 
structure of the planets rely in turn on assumptions about the composition
and temperature profiles through the planet.

Theories of planetary formation can be investigated by 
comparing the structure of exoplanets with those within the solar system.
Another motivation is to estimate the occurrence of Earth-like planets,
in terms of mass and composition, and also those that might be habitable for
life.

Deductions about planetary structures, i.e. the composition profile,
depend on the compressibility of the possible compositions 
thought to occur.
The compressibility is needed over the range of pressures and temperatures
occurring within each planet.  
The compressibility is found from the derivative of the pressure-density relation
at the appropriate temperature, which can be determined from the
equation of state (EOS) for the particular composition of matter
of interest.
The development of EOS has been driven most by research in weapons
(explosives and projectile impacts) \citep[for instance][]{shock},
geophysics
\citep[e.g.][]{Ahrens1964,Ahrens1966,Alfe2001,Stacey2004,Stixrude2009},
and inertial confinement fusion \citep{ICF}.
There is a perception that experimental and theoretical methods for
determining EOS are not available in regimes necessary to understand
the internal structure of planets for pressures between 200\,GPa
and 10\,TPa, i.e. from the limit of diamond anvil data to the onset of the
Thomas-Fermi-Dirac (TFD) regime \citep{Seager2007,Grasset2009}.
Some studies \citep[e.g.][]{Seager2007} have considered sub-TFD EOS
with care, but it is common practice even when accurate theoretical calculations
are available to represent the material with {\it ad hoc} functional forms,
potentially leading to problems when extrapolating beyond the bounds of the
constraining data and in switching over to TFD at high pressures.

Although there is a definite need for more theoretical and experimental studies,
appropriate theoretical techniques are just as accurate above 200\,GPa as below,
and are more than adequate to draw inferences about the structure of exoplanets;
moreover, shock and ramp loading experiments can readily explore
states up to $\sim$5\,TPa and are suitable to test and calibrate EOS.
In this paper, we study the validity of electronic structure methods
for predicting EOS up to the $o(100)$\,TPa pressures
\footnote{A pressure of 1\,TPa is 10 million atmospheres.}
apparently occurring in exoplanets,
and the capability of dynamic loading experiments to measure relevant states.
We calculate mass-radius relations for several compositions of matter
representing different classes of, and layers in, planets, discussing
the validity of the EOS used.
Finally, we compare the mass-radius relations with representative
planets and exoplanets.

\section{Matter at high pressure}
The quasistatic structure of self-gravitating bodies depends on the
scalar EOS, which can be described by an appropriate free energy expressed
as a function of its natural variables, such as the Helmholtz free energy
$f(\rho,T)$, where $\rho$ is mass density and $T$ temperature.
In principle, one could consider the composition as a set of explicit
parameters in the EOS.
In practice, planetary structures are thought to comprise layers
in each of which a single composition, or range of compositions, 
dominates the EOS, such as Fe in the core of rocky planets.
Therefore, we consider a separate EOS for each layer.

As with dynamic loading situations, 
the pressure $p\equiv\partial f/\partial v|_T$ is the most
directly useful aspect of the EOS for calculating and interpreting
planetary structures.
Almost invariably, a thermodynamically incomplete EOS is used:
$p(\rho,T)$ or $p(\rho,e)$ where $e$ is the specific internal energy.
Planetary structures may be constrained to follow a specified
temperature profile, in which $p(\rho,T)$ is the more convenient form,
or an isentrope, for which $p(\rho,e)$ is convenient since
$p=-\partial e/\partial v|_s$ where $v=1/\rho$.

In planets, brown dwarfs, and main sequence stars, 
the EOS is dominated by electrostatic
forces and Pauli exclusion among the electrons and ions, 
rather than by strong-force interactions between the nuclei.
In stars, the radiation pressure must be included, and the entropy
is much higher, exploring a different region of the EOS.

\subsection{Theory}
In this section, we describe key theoretical methods and approximations
used when predicting EOS,
and comment on their applicability to states likely to occur in exoplanets.

The relevant EOS of matter can be calculated using electronic structure 
theory. 
For a representative set of atoms defining the composition,
thermodynamic potentials can be calculated as a function of $\rho$
(i.e. system volume) and $T$.
Because the mass of the electrons is so much less than that of the nuclei,
the state of the electrons can generally be considered with respect to
the instantaneous positions of the nuclei -- the Born-Oppenheimer approximation.
Forces on the nuclei can also be calculated with respect to their positions and
the distribution of electrons.
The energy of the system can be thought of as comprising the 
ground state energy for stationary ions, thermal motion of the ions,
and thermal excitation of the electrons out of their ground state.
In principle, all contributions should be calculated self-consistently.
However, for EOS, it is generally a good approximation to partition
the free energy into the cold compression
curve, thermal motion of the ions, and thermal excitation of the electrons,
\begin{equation}
f(\rho,T)=f_c(\rho)+f_i(\rho,T)+f_e(\rho,T)
\end{equation}
\citep[see][]{Swift_SiEOS_2001}.
To describe the material state and the dynamics of most atoms,
the quantum nature of the nuclei can be ignored, allowing their treatment as
point-like masses exhibiting Newtonian dynamics.
The exception is H, for which the quantum behavior of the proton can be
represented as a correction to the potential experienced by the electrons.

The motion of the nuclei can be calculated with respect to time.
This technique, known as first-principles molecular dynamics (FPMD)
or quantum molecular dynamics (QMD)
\citep{Michielsen1996},
is particularly appropriate
for fluids, unknown or ill-defined crystal structures, and multi-species
compositions of matter where it is not clear {\it a priori} where particular
species will be located.
In addition to the EOS, FPMD has been used to predict transport properties, 
including the electric and thermal conductivities
\citep{Recoules2009,Hamel2011,Holst2011}
and viscosity \citep{Alfe2000a,Clerouin2002},
which are important in understanding the formation and thermal profile of planets
as well as the generation of their magnetic fields.
FPMD has also been used in studies of the free energy profiles of mixtures
to determine possible miscibility gaps which could have a large impact on 
planetary evolution and structure \citep{Morales2009}. 

For crystalline structures, the motion of the nuclei can be calculated
in terms of oscillations about their equilibrium positions,
i.e. phonon modes.
In principle, phonon modes should be calculated self-consistently
with electron excitations;
in practice, for most conditions encountered in planets, the ion-thermal 
motion can be calculated from the phonon density of states at $T=0$,
\footnote{All temperatures here are defined with respect to absolute zero.}
and the electron-thermal contribution can often be ignored.
$f_c$ comprises the ground state energy of the system as a function
of compression, $e_c(\rho)$, and any configurational entropy
that may be associated with different structural polymorphs or the entropy
of mixing in an impure system
\citep{Kittel1980}.
One may calculate $e_c$ by setting up a configuration of nuclei,
and finding the ground state of the Hamiltonian of the electrons.
The configuration of the nuclei may be altered or relaxed under the net
force they experience to find the lowest-energy structure,
but they are often held fixed in likely structures.
For well-defined crystalline structures,
phonon modes can be predicted by calculating the electrostatic forces
on the nuclei as each is displaced from equilibrium.
Explicit calculation of the phonons is desirable below the Debye temperature,
where the zero-point motion of the nuclei and the freezing out of modes may
have a significant effect on the EOS \citep{Swift_SiEOS_2001}.

Thermal excitation of the electrons can be calculated from the band structure,
which is the set of eigenstates of the electronic Hamiltonian.
For sufficiently high temperatures, the eigenstates must be calculated
self-consistently with their population, but accurate calculations can be made
to $\sim$eV temperatures using the $T=0$ band structure. 

The key physics, and thus the source of limitations in the accuracy and
validity of the EOS, is the electronic structure calculation.
The challenge is in representing the multi-fermion nature
of the electrons accurately, but with a method that is computationally
tractable for real materials.
Path integral methods, such as path-integral Monte-Carlo (PIMC),
can be used for direct, rigorous calculations of multi-particle states
\citep{Militzer2009}.
However, these methods are practicable mostly for low-$Z$ materials,
and have not been applied systematically (or, generally, at all) 
to compositions of matter relevant to ice and rocky planets.
There is, however, no reason to suspect that these techniques are unsuitable
under exoplanet conditions {\it per se}.

Although TFD-based electronic structure theory \citep{Salpeter1967} 
is often regarded as an adequate treatment for pressures
over $\sim$10\,TPa \citep[for instance][]{Seager2007,Grasset2009}, 
it does not capture the effects of electronic shell structure
which are expected at high pressures \citep{Liberman1979}.
However, shell effects such as pressure ionization are captured by
the electronic structure techniques considered here.

Most theoretical EOS calculations use density functional theory (DFT)
\citep{Hohenberg1964,Kohn1965,Perdew1992,Perdew1994} 
and its variants to treat exchange and correlation between the electrons
via functionals of the electron density, calibrated to reproduce 
the same system energy as techniques that treat the electron wavefunctions
more directly.
DFT functionals are calibrated against calculations of idealized
electron gases, which may be performed up to arbitrarily high densities:
they should be no less accurate for the relatively modest absolute compressions 
occurring in the cores of exoplanets than at $p=0$.
DFT calculations typically reproduce the observed mass density 
to within a few percent \citep[for instance][]{Swift_SiEOS_2001,Ackland2002}, 
which is an observable discrepancy
when compared with dynamic or quasistatic loading measurements.
Calculations can be corrected by adjusting the energy to reproduce the
mass density at zero pressure, which is most accurately measured.
The resulting EOS -- which we have termed `ab fere initio' to distinguish
them from uncorrected ab initio EOS --
then typically reproduce high pressure measurements to within their
uncertainties \citep{Swift_SiEOS_2001,Swift_NiAlEOS_2007}.

In many situations, electrons closer to the nuclei are not affected by
changes in the compression or temperature,
and the states of the other electrons may be calculated more efficiently
by subsuming the inner electrons into a modified nuclear potential:
the pseudopotential method \citep{Payne1992}.
Pseudopotential calculations can become inaccurate when the nuclei
are compressed sufficiently closely together.
The validity can be checked by comparing against all-electron calculations,
and may be accurate to severalfold compressions or pressures of terapascals
\citep{Morales2009}.

With these caveats, pseudopotential techniques and the DFT construct in general are not inherently
unsuitable for predicting EOS in the planetary structure regime.
However, detailed calculations of many-species compounds are computationally
expensive, and predictions of polymorphic structures are sensitive to
relatively small inaccuracies in the computational methods.
It is highly desirable to compare EOS predictions against experimental
measurements.

For the most part, we have used previously-developed EOS based on experimental
data or validated electronic structure calculations up to the limit of
available data, and blending into TFD predictions at high compressions
\citep{SESAME}.
These EOS incorporate thermal effects, allowing us to investigate the
sensitivity of planetary structures to temperature profile.
For Fe and Fe-Ni, we have compared predictions from such EOS with 
3D electronic structure calculations extended to higher pressures than
have been reported previously.
Previous theoretical studies include very careful treatments of the EOS of Fe
up to pressures and temperatures representative of the
Earth's core
\citep[in particular][and references therein]{Wasserman1996,Stixrude1995,Alfe2000,Sola2009,Belonoshko2008},
and it is very desirable to perform equivalent studies to higher pressures
and temperatures necessary for the study of exoplanets.
For our present purposes, it is most useful to compare EOS for different
compositions of matter constructed according to consistent prescriptions.
The use of wide-ranging EOS constructed using optimized algorithms,
and wide-ranging sets of electronic structure calculations made using a 
consistent method that reproduces TFD at extreme compressions,
allows us to avoid any reliance on extrapolating using {\it ad hoc} 
functional forms such as
Vinet and Birch-Murnaghan for pressure-density relations, which can give
unquantified uncertainties outside the range of the fitting data.

\subsection{Dynamic loading experiments}
The canonical experimental technique for studying the properties at matter
at high pressure is shock loading, using a variety of methods to
induce a shock wave.
Shock measurements of EOS are often performed relative to 
a reference material, but
an attraction of shock loading is that experiments can in principle be
configured to yield absolute measurements, if the shock is induced by
the impact of a projectile with a target of the same material.
Indeed, the pressure standards in static compression apparatus
such as diamond anvil cells
are ultimately calibrated against absolute shock measurements.

Although the time scales in dynamic loading experiments are typically
nanoseconds to microseconds,
typical equilibration times for electrons and atomic vibrations 
are much shorter, so inferred states used for testing and 
calibrating EOS are in thermodynamic equilibrium and thus equivalent
to quasistatic compression measurements made in presses such as diamond anvil 
cells.
Indeed, the difference in time scale between dynamic and quasistatic loading
is less than the difference between the latter and planetary ages.
Although EOS measurements are in thermodynamic equilibrium with respect
to a given phase of matter, the time dependence of phase transitions must
be considered: phase changes often occur with a significant degree of
superpressurization or superheating in dynamic loading experiments and indeed
in quasistatic loading, compared with the equilibrium phase boundary.
The effects of time dependence are also evident as hysteresis in the location
of the phase change on loading compared with unloading.

In shock loading, the entropy increases with compression,
so the temperature rises faster with compression than it does 
along an isentrope.
The Gr\"uneisen parameter and heat capacity of a material, which may be
estimated theoretically or experimentally, can be used to predict the EOS
away from the locus of shock states.
Measurements along the principal shock Hugoniot have been used in this way 
for many years to estimate the principal isentrope and the cold curve
\citep{Bushman1993}.
While a direct measurement of a relevant state is preferable,
shock-derived EOS are likely to be adequate for exoplanets over a wide
range of pressures.
Furthermore, if a theoretical EOS is validated by shock experiments,
this provides reasonable confidence that the EOS is valid at lower
temperatures.
A more serious limitation with shock experiments is that, for a given
material and starting state, there is a limit to the compression that can be
achieved by the passage of a single shock, and therefore a limit
to the range of compressions that can be deduced along isentropes.
However, this compression (discussed further below for different
compositions of matter) is adequate to explore matter into the massive
exoplanet regime.
Shock measurements have been made to pressures of up to 10\,TPa
using radiation from underground nuclear explosions \citep{Ragan1984}. 

The compression limit can be circumvented through the use of multiple shocks,
which induce lower temperatures than a single shock to the same pressure.
The ultimate limit for an infinite sequence of infinitesimal shocks is
a ramp compression.
Ramp compression of condensed matter was first demonstrated using the 
expanding reaction products from a chemical explosive \citep{Barnes1974}, 
and has 
subsequently been demonstrated using pulsed magnetic fields 
at the Z accelerator \citep{Reisman2001}
and high power lasers.
With lasers, the pulse energy may be used to vaporize a foil which
then loads the sample in an analogous way to the chemical reaction
products \citep{Edwards2004},
the pulse may ablate the sample directly with an intensity history
designed to induce a ramp in pressure \citep{Swift_lice_2005},
or the sample may be driven from a hohlraum with a power history
chosen to control the hohlraum temperature history such that a ramp
is induced.
Ramp loading using the last variant has been demonstrated to
$\sim$1\,TPa on the OMEGA laser \citep{Bradley2009}, 
with near-term experiments on Fe planned to 
reach 2\,TPa at the National Ignition Facility (NIF).
Experimental techniques have thus been available to develop and test EOS
under conditions relevant to exoplanets,
and planned developments should provide the first direct measurements of
matter under exoplanet core conditions.

\section{Mass-radius relations}
If the rotation rate is not extreme,
self-gravitating bodies are close to spherical.
Spherical structures are certainly an adequate starting place for studies
of exoplanets.
The condition for isostatic equilibrium is that the
stress induced by pressure variations is balanced by the gravitational
acceleration $g(r)$:
\begin{equation}
\grad\,p(r) = -\rho(r)g(r).
\label{eq:gradp}
\end{equation}
For Newtonian gravitation, by Gauss' theorem,
$g(r)$ can be expressed in terms of the mass enclosed within a given radius
$m(r)$:
\begin{equation}
g(r) = \frac{Gm(r)}{r^2}
\label{eq:gr}
\end{equation}
so
\begin{equation}
\diffl{g(r)}r = G\left(\frac 1{r^2}\diffl{m(r)}r-2\frac{m(r)}{r^3}\right).
\label{eq:dgdr}
\end{equation}
$m(r)$ can be calculated simply from the distribution of mass density,
\begin{equation}
m(r) = 4\pi\int_0^r r'^2\rho(r')\,dr'
\label{eq:mr}
\end{equation}
or
\begin{equation}
\diffl{m(r)}r = 4\pi r^2\rho(r).
\label{eq:dmdr}
\end{equation}

When the total mass $M$, outer radius $R$, and surface composition 
are known with reasonable confidence,
and the objective is to determine an aspect of the internal structure 
such as the core radius,
the equations can be integrated from the surface inward,
starting at the known $p(R)=0$ (hence $\rho(R)$) and $g(R)=GM/R^2$
\citep{Swift_planetstruc_2009}.
The internal structure parameter is found as the solution of a shooting
problem to ensure that $m(r)\rightarrow 0$ as $r\rightarrow 0$.
To calculate generic mass-radius relations for a uniform composition,
it is most efficient to start at the center,
where $g(0)=0$ and $m(0)=0$,
choose the core pressure $p_c=p(0)$,
and integrate outward until $p=0$.
In this way, the mass-radius condition is obtained in terms of
$p_c$ as an independent parameter: $\{R,M\}(p_c)$.

These solution methods are unwieldy when it is desired to constrain the
overall composition as a mixture of components, such as a fixed ratio
between Fe, rock, and H/He.
In this case, the isostatic structure may be found more conveniently by
starting with the desired masses of each component, and solving the 
system of equations as an elliptic problem.
A numerical solution may be found iteratively by taking an approximate
solution (which may initially be concentric shells of material at their
$p=0$ density) and improving it by calculating the net stress gradient at
each radius
\begin{equation}
\grad\,\tau=\grad\,p+\rho g
\end{equation}
and
\begin{equation}
\pdiffl pr\simeq \pdiffl p\rho\pdiffl\rho r
\end{equation}
(ignoring the relatively slow change of $g$ with deformation,
which is valid near equilibrium),
and hence estimating a change in radius for each shell such that the
net pressure is zero.
The radial variation was represented discretely by values at a finite
series of radii.
Multiple iterations are necessary to allow the effect of a net pressure
at some radius to be balanced through equilibration of the whole structure,
but standard techniques may be used for solving elliptic equations efficiently,
such as successive over-relaxation \citep{Press1989}.
A series of solutions describing a mass-radius relation
can be found efficiently by working upward in mass,
and increasing the density at each radius by a constant factor,
then calculating the modified equilibrium structure.
We verified that this method of solution gave the same result as 
integration from the center to the surface, by comparison with 
structures comprising pure Fe.

To close the system of equations, a $p(\rho)$ relation is needed for the
material.
The choice of $p(\rho)$ relation implies a choice for the entropy or
temperature profile of the body.
In the present study we usually chose
an isentrope passing through a reasonable surface state, such as STP
for rock or metal, or a few atmospheres pressure and cryogenic temperatures
for compositions that are gaseous at STP.
Isentropes were calculated by numerical integration of the relation
\begin{equation}
\left.\pdiffl ev\right|_s=-p(\rho,e)
\end{equation}
using a procedure valid for EOS of arbitrary form \citep{Swift_genscalar_2008}.
For a few sample compositions, structures were calculated for $p(\rho)$
along the cold curve.
The precision of $p(\rho)$ curves necessary to be useful for exoplanet
structures is lower than for solar system planets, 
and the difference in structures
between the STP isentrope and cold curve was found to be negligible over the 
range of pressures considered, as shown below.

Mass-radius relations were calculated for a series of compositions of matter
relevant to (exo)planets,
using as a baseline EOS developed from and for shock wave applications,
in particular the SESAME library \citep{SESAME}, for which the constituent
assumptions and calibrations are reasonably well documented.
SESAME EOS are tabulated functions $\{p,e\}(\rho,T)$.
Most SESAME EOS are tabulated over a wide enough range of states to be valid
(at least in principle, to some finite precision)
for massive exoplanets and star formation. They were constructed to
incorporate experimental data, mainly shock measurements, where available.
Experimental measurements not used in the construction of the EOS, including
shock states, release from shock, and ramp compression, are
frequently compared against the SESAME EOS, so there is a relatively rich
literature on experimental validation. 
Many of the EOS were constructed using the same theoretical and
modeling approaches, so conclusions drawn from the use of one EOS are more
likely to apply to other materials.
When these EOS are compared with subsequent shock or ramp measurements
at higher pressures, it is quite usual to find significant differences, i.e.
inaccuracies in the EOS, but they are usually at the level of a few times
the experimental uncertainty \citep[for instance][]{Hicks2005,Knudson2009}, 
which would not have much effect on the mass-radius relation.

For all undifferentiated compositions, the low-mass limit behaves as 
\begin{equation}
M = \frac 43 \pi R^3 \rho_0
\end{equation}
and deviates as matter within the planet is compressed gravitationally 
above its zero-pressure density $\rho_0$.
The deviations become substantial as the central pressure 
approaches the bulk modulus.
As the mass increases further, an increasing proportion is compressed
to significantly higher density, until eventually any increase in mass is
accommodated entirely by density increase, leading to a maximum in radius.

\subsection{Fe and Fe-Ni}
\begin{figure}
\begin{center}\includegraphics[width=\textwidth]{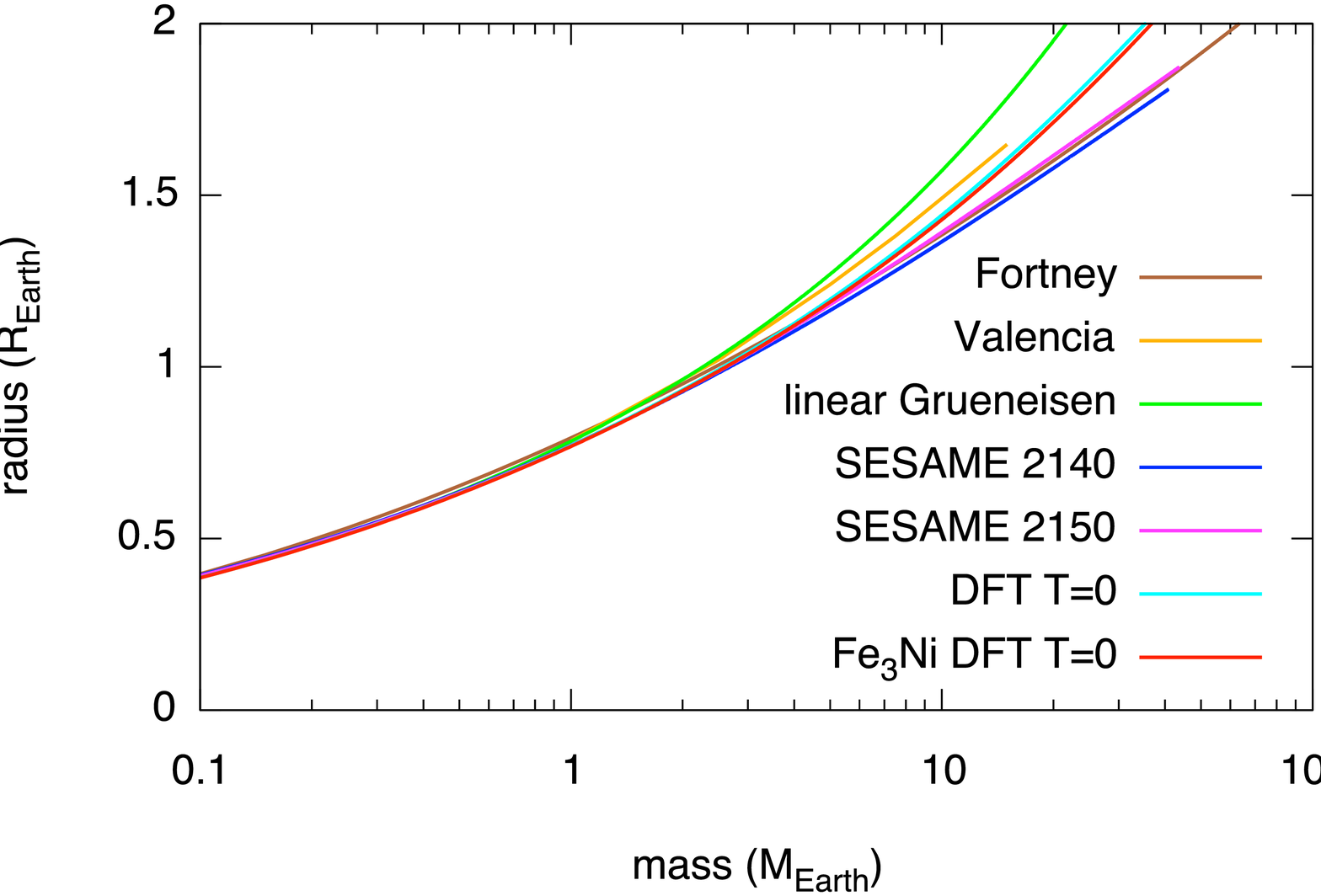}\end{center}
\caption{Mass-radius relations calculated for different Fe EOS
   and for Fe$_3$Ni.}
\label{fig:Femr}
\end{figure}

\begin{figure}
\begin{center}\includegraphics[width=\textwidth]{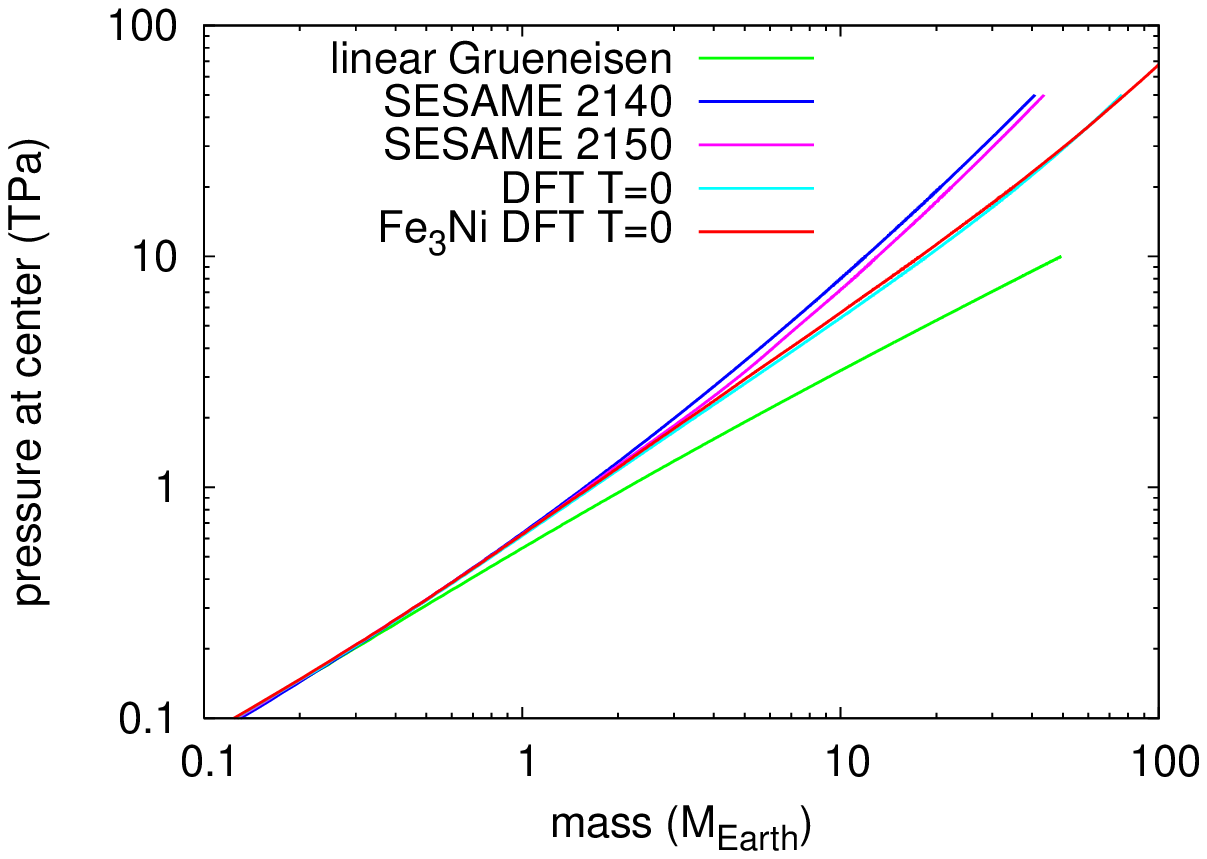}\end{center}
\caption{Predicted variation of core pressure with mass for different Fe EOS
   and for Fe$_3$Ni.}
\label{fig:Femp}
\end{figure}

For Fe, which is the dominant component of interest at the highest pressures,
we are developing EOS better optimized to exoplanet core states,
to complement NIF experiments \citep{Swift_Feeos_2010}.
Thermodynamically complete versions of these EOS were not available for this
study: only the cold curves.
The effect of temperature was assessed by comparing structures calculated with
$p(\rho)$ along an isentrope compared with the cold curve, for 
thermodynamically complete Fe EOS, and was found to be negligible,
in agreement with previous work using a simplified thermal contribution
\citep{Seager2007}.
The mass-radius relation derived from
our new Fe cold curve, which was calculated with DFT using the local density 
approximation and Troullier-Martins pseudopotentials \citep{Troullier1991}, 
agreed well with the relations from SESAME EOS 2140 \citep{BarnesRood1973} 
and 2150 \citep{Kerley1993}
(Fig.~\ref{fig:Femr}).
Fe$_3$Ni was considered as a representative Fe-Ni composition,
calculated using the same DFT method as for previous studies
of intermetallic compounds \citep{Swift_NiTi_2005,Swift_NiAlEOS_2007}.
The pseudopotential for Ni was used previously for constructing EOS
for NiAl \citep{Swift_NiAlEOS_2007}.
Comparing $p(\rho)$ relations, Fe$_3$Ni was significantly softer than Fe
or a segregated Fe-Ni mixture of the same composition,
but the mass-radius relations did not differ significantly.
The mass-radius relation was also calculated for the principal isentrope
from the linear Gr\"uneisen EOS
found previously to perform well for rocky planets 
\citep{Swift_planetstruc_2009};
it deviated from the other Fe relations for masses above approximately
twice that of the Earth (\ME),
demonstrating the inaccuracy of mass-radius relations deduced from
EOS extrapolated from low-pressure measurements.
Mass-radius relations deduced from the DFT cold curves were very similar to
those from the SESAME EOS up to around 1\,\RE\ or 3\,\ME\ ($\sim$1\,TPa),
above which the properties calculated with DFT became progressively softer.
This observation highlights the importance of ramp-loading experiments 
to several terapascals. 
With the exception of the linear Gr\"uneisen EOS, these relations
are similar, but not identical, to relations used recently in the 
interpretation of exoplanet CoRoT-7b \citep{Leger2009}
and for Fe mass-radius relations \citep{Fortney2007,Valencia2010}.
Interestingly, the relation by Valencia et al follows the linear Gr\"uneisen curve up to
around 2\,\ME, then approaches the $T=0$ DFT relations above 10\,\ME;
the Fortney et al relation is stiffer (larger radius) up to around 1\,\ME,
then follows the relation for SESAME 2150 very closely.
The mass-radius relations derived from the SESAME EOS were denser above 1\,\ME.
(Figs~\ref{fig:Femr} and \ref{fig:Femp}.)

For Fe, the maximum shock compression deduced from the SESAME EOS
was a factor of around 5.2, calculated using a solution method
valid for general forms of the EOS \citep{Swift_genscalar_2008}.
At the same mass density on the principal isentrope, the pressure is
approximately 11\,TPa.
Thus single shock experiments, corrected appropriately for thermal effects,
can plausibly explore planetary pressures to this regime.
Impact-induced shock experiments, with the projectile driven by
gas, chemical propellant, or high explosive, have been performed
on Fe to around 1\,TPa by several groups \citep{vanThiel1966}.
At higher pressures, data are very rare.
Nuclear-driven shock impedance mismatch measurements \citep{Ragan1984}
reached a pressure which we estimate to be $4.57\pm 0.26$\,TPa.
Further measurements above 1\,TPa are desirable.

\subsection{Rocks}
\begin{figure}
\begin{center}\includegraphics[width=\textwidth]{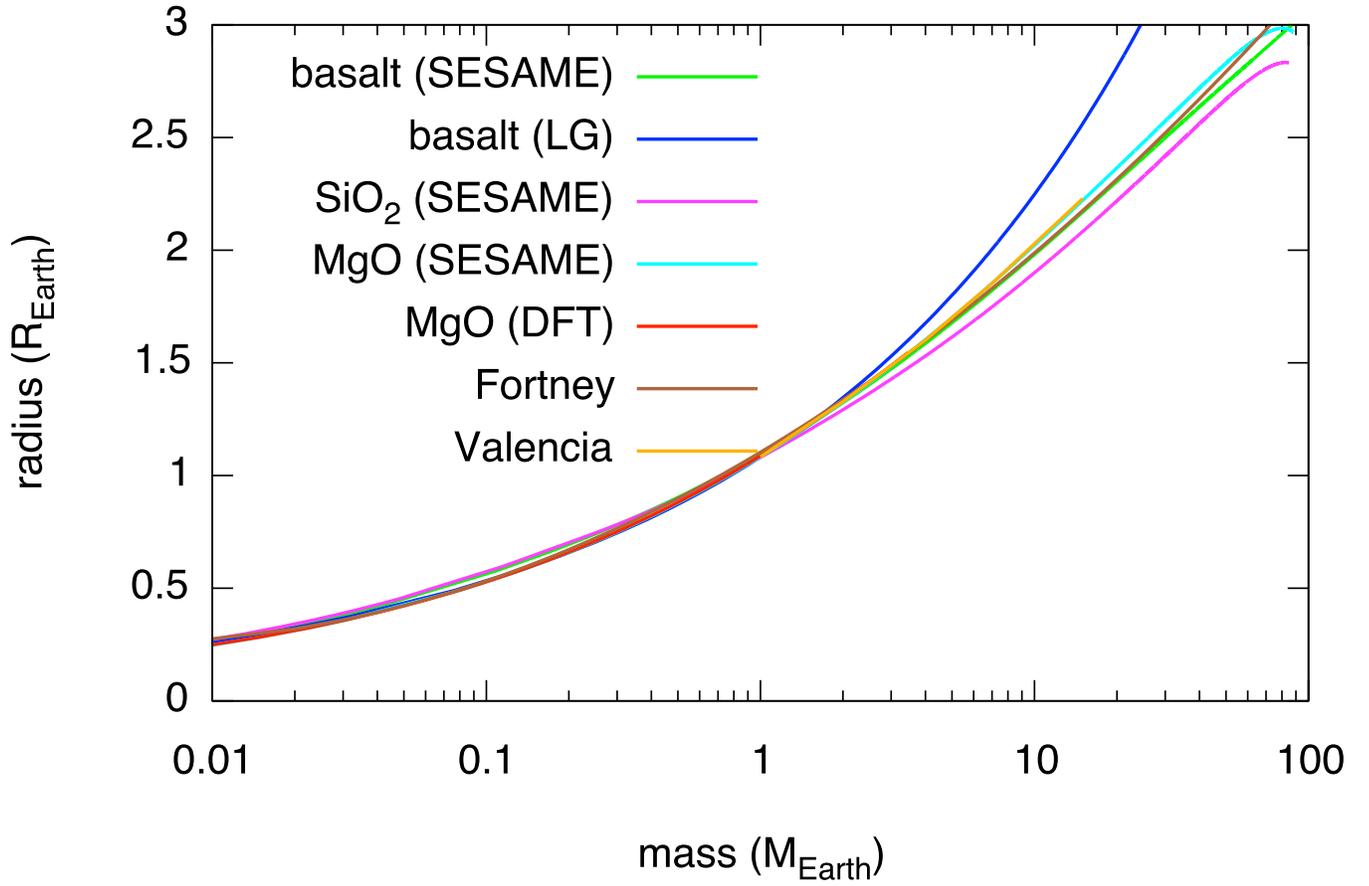}\end{center}
\caption{Mass-radius relations calculated for SiO$_2$, MgO, and basalt,
   including a simplified, linear Gr\"uneisen EOS (LG),
   and comparing with previous calculations \citep{Fortney2007,Valencia2010}.}
\label{fig:rockmr}
\end{figure}

\begin{figure}
\begin{center}\includegraphics[width=\textwidth]{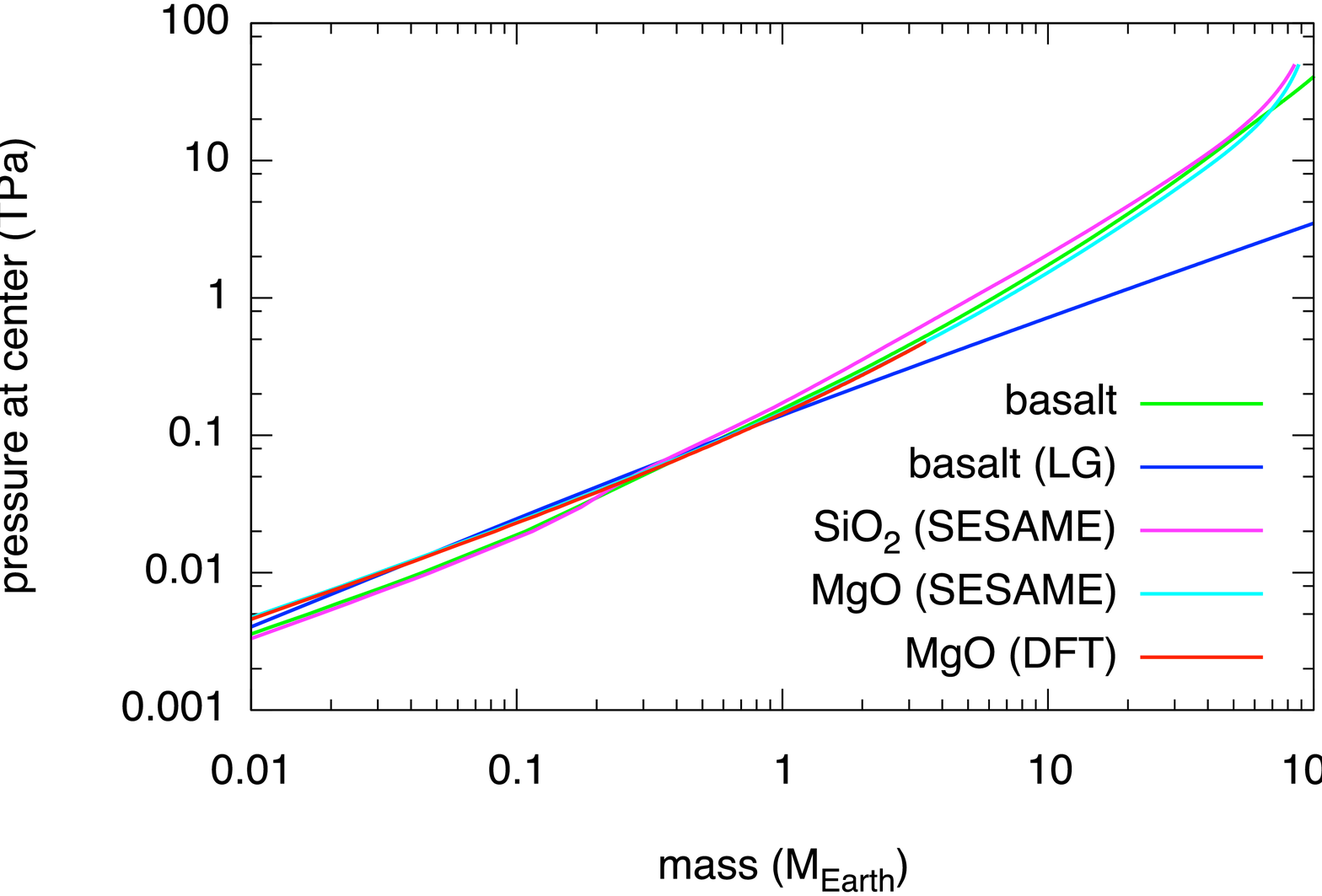}\end{center}
\caption{Predicted variation of core pressure with mass for SiO$_2$, MgO, and basalt,
   including a simplified, linear Gr\"uneisen EOS (LG).}
\label{fig:rockmp}
\end{figure}

High quality theoretical and experimental studies have been performed for
mantle constituents such as MgSiO$_3$ and SiO$_2$ up to pressures 
around 150\,GPa relevant to Earth's core-mantle boundary,
and the effect of Fe ions
\citep[see][]{Drummond2002,Caracas2008,Stixrude2010,Wentzcovitch2010,Stixrude2011}.
As compositions representative of the variations in rocks,
mass-radius relations were calculated for SiO$_2$, MgO, and basalt.
SiO$_2$ was represented by SESAME EOS 7383 \citep{JohnsonLyon1984}.
MgO was represented by SESAME EOS 7460 \citep{BarnesLyon1988_MgO}, 
and also by a DFT-derived two phase EOS including quasiharmonic phonons
\citep{Drummond2002,Luo_MgO_2004}, but extending only to $\sim$0.4\,TPa.
The mass-radius relations for the different MgO models were indistinguishable,
demonstrating insensitivity of the mass-radius relation to the inclusion
of phase transitions, for current exoplanet applications.
Basalt was represented by SESAME EOS 7530 \citep{BarnesLyon1988_basalt}.
`Basalt' refers to the composition
(by number of atoms: 60.11\%\ O, 18.26\%\ Si, 5.96\%\ Al, 4.01\%\ Ca,
3.48\%\ Fe, 3.39\%\ Mg, 2.18\%\ H, 1.56\%\ Na, 0.55\%\ Ti, 0.38\%\ K,
0.07\%\ P, 0.06\%\ Mn)
and initial state: the EOS was derived from
shock data including phase changes, and was blended into 
TFD predictions at high pressures.
We also compared the linear Gr\"uneisen EOS for basalt found previously
to give better than expected structures for the rocky planets
\citep{Swift_planetstruc_2009};
its mass-radius relation was similar to that of the SESAME EOS to several
tenths of a terapascal, and then diverged significantly.
The mass-radius and pressure-radius relations were similar but distinguishable 
on scales extending to exoplanets, giving an indication of the possible
uncertainty in inferences about rocky planet structure in the absence
of constraints on the composition:
the mass-radius relation does show sensitivity to the composition,
but the sensitivity is small compared to current uncertainties in
{\it exoplanet} observables.
Recent mass-radius calculations by \cite{Valencia2010} for Mg-silicate planets 
were
almost indistinguishable from the MgO relation; slightly older calculations
by \cite{Fortney2007} using a linear mixture of 38\%\ SiO$_2$,
25\%\ MgO, 25\%\ FeS, and 12\%\ FeO
followed our calculation for basalt (SESAME) very closely
to several tens of \ME.
(Figs~\ref{fig:rockmr} and \ref{fig:rockmp}.)

\subsection{Ices}
{\it Ab initio} treatments of mixtures of molecules of low molecular weight,
usually called ices (regardless of being in the solid or fluid state), are complicated and computationally expensive 
because of the possibility of segregation of species and the formation of
a wide variety of chemical species. At issue here is the uncertainty of the equilibrium concentration of the different molecules.
Such calculations have been performed for relatively small numbers of
thermodynamic states \citep{Chau2011_ice}, and will in future allow wide-ranging
EOS to be predicted as a function of composition.
A complementary approach is to calculate the equilibrium chemical composition
given the thermodynamic state and atomic composition, using measured or
estimated thermodynamic potentials for each possible chemical specie.
Chemical equilibrium calculations have been used to predict the EOS of the
reaction products of chemical explosives up to temperatures of 
several hundred gigapascals and temperatures of $\sim$eV, including many
species in common with planetary ices \citep{CHEETAH,MulfordSwift2006}.
However, the thermodynamic potentials are not necessarily well-constrained for
the hot, compressed states occurring within exoplanets.

\begin{figure}
\begin{center}\includegraphics[width=\textwidth]{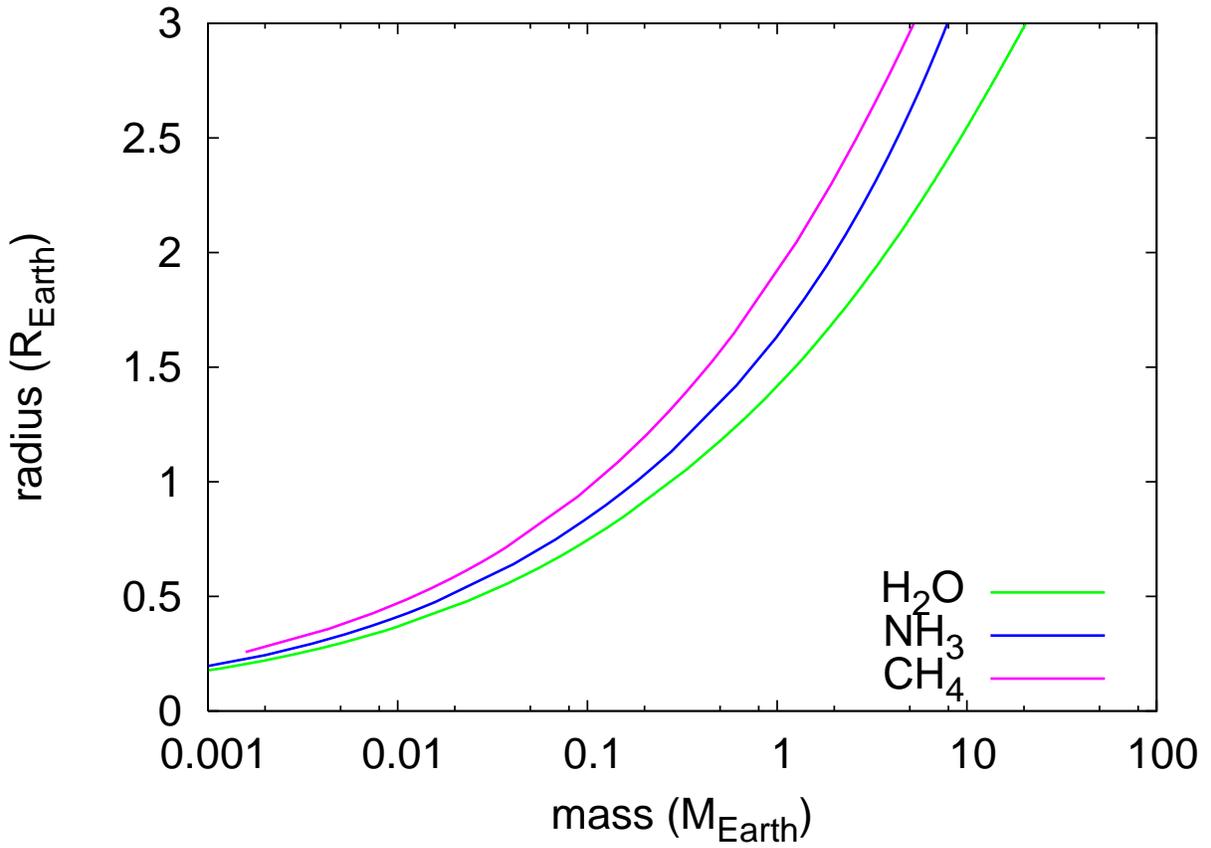}\end{center}
\caption{Mass-radius relations calculated for H$_2$O, NH$_3$, and CH$_4$ ices.}
\label{fig:icemr}
\end{figure}

\begin{figure}
\begin{center}\includegraphics[width=\textwidth]{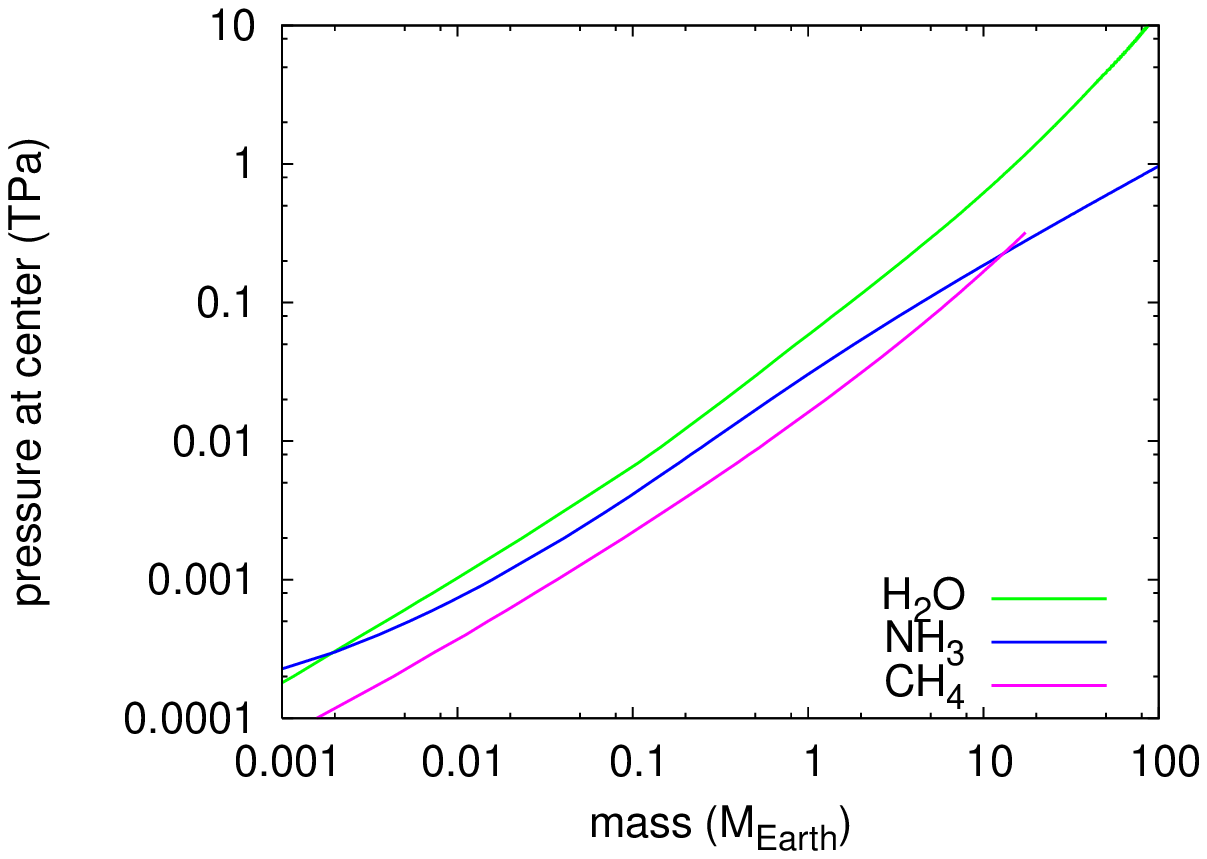}\end{center}
\caption{Predicted variation of core pressure with mass for H$_2$O, NH$_3$, and CH$_4$ ices.
   The crossing of the CH$_4$ by the NH$_3$ curve is at a mass density
   much greater than the shock data used to construct the NH$_3$ EOS,
   and should not be regarded as a meaningful prediction.
   The crossing of the H$_2$O and NH$_3$ curves is in a regime where the
   EOS should be adequately accurate.}
\label{fig:icemp}
\end{figure}

As compositions representative of the variations in ice giants,
mass-radius relations were calculated for H$_2$O, NH$_3$, and CH$_4$,
for isentropes chosen to pass through a condensed phase so that
the planet surface could be defined as zero pressure.
The calculated radius is thus a lower bound for
planets with a thick atmosphere, as is likely to be the case for those
orbiting close to their star.
Several EOS were available for H$_2$O; SESAME EOS 7154 was the most
appropriate for these pressures \citep{JohnsonLyon1990}.
The STP isentrope was used.
The only SESAME EOS available for NH$_3$ was 5520 \citep{Johnson1982_5520},
fitted to National Bureau of Standards (NBS) gas phase data,
and was tabulated to a maximum mass density of 0.765\,\SIdens,
which is barely greater than typical for liquid ammonia.
An empirical linear Gr\"uneisen EOS was constructed 
from shock wave data on liquid ammonia
at an initial temperature of 203\,K \citep{Swift2011_NH3EOS},
using an approximate treatment for off-Hugoniot states.
The corresponding mass-radius calculation should be regarded with caution as
it is not constrained by experimental data for pressures above a few tens
of gigapascals.
The mass-radius relation and central pressure were in reasonable agreement
with predictions using SESAME EOS 5520 up to a mass $\sim$0.1\,$M_E$
above which the isentrope from the tabular EOS became unusable.
For CH$_4$, three SESAME EOS were available:
5500 and 5501 \citep{Kerley1980}, both tabulated to 2.5\,\SIdens\ and 
respectively with a Maxwell construction and van der Waals loops in the
liquid-vapor region;
and 5502 \citep{Johnson1984_5502}, tabulated to 0.47\,\SIdens\ and based on NBS
gas phase data. 
The mass-radius relation was calculated for 5500, which is the most relevant
for planetary structures, for an isentrope passing through 70\,K 
at zero pressure.
The difference between H$_2$O and CH$_4$ was considerable:
it is difficult to infer a composition for an ice giant from the mass-radius
measurement alone to a factor $\sim$2 in mass or $\sim$25\%\ in radius,
without further constraints on the composition.
(Figs~\ref{fig:icemr} and \ref{fig:icemp}.)

\subsection{Hydrogen}
Jovian planets are thought to be composed predominantly of a mixture of
H with approximately 5-10\%\ He by number of atoms.

H is theoretically interesting because quantum mechanical uncertainty in the
position of the proton is more important than for other elements,
and may affect the EOS.
Significant work has been devoted to predicting the EOS
of H and H/He mixtures \citep{Saumon1992,Delaney2006,Morales2009}.
These H/He mixtures are challenging because even the relatively small amount of helium involved is thought to greatly impact the EOS. 
Specifically, the insulator-to-metal transition that occurs at $o(100)$\,GPa
pressures in pure H \citep{Morales2010,Lorenzen2010} is thought to be 
suppressed to even higher pressures and temperatures with the inclusion of
trace amounts of He \citep{Hensel1999}. 
Furthermore, helium is thought to segregate
 (i.e. form droplets) in the metallic region of the planet\citep{Morales2009}, where the resulting He rain is thought to produce heat as the 
 droplets fall down the gravitational well. 
This effect could have a disproportionately large effect on the mean EOS and on the
differentiation of a planet by gravitational separation.

\begin{figure}
\begin{center}\includegraphics[width=\textwidth]{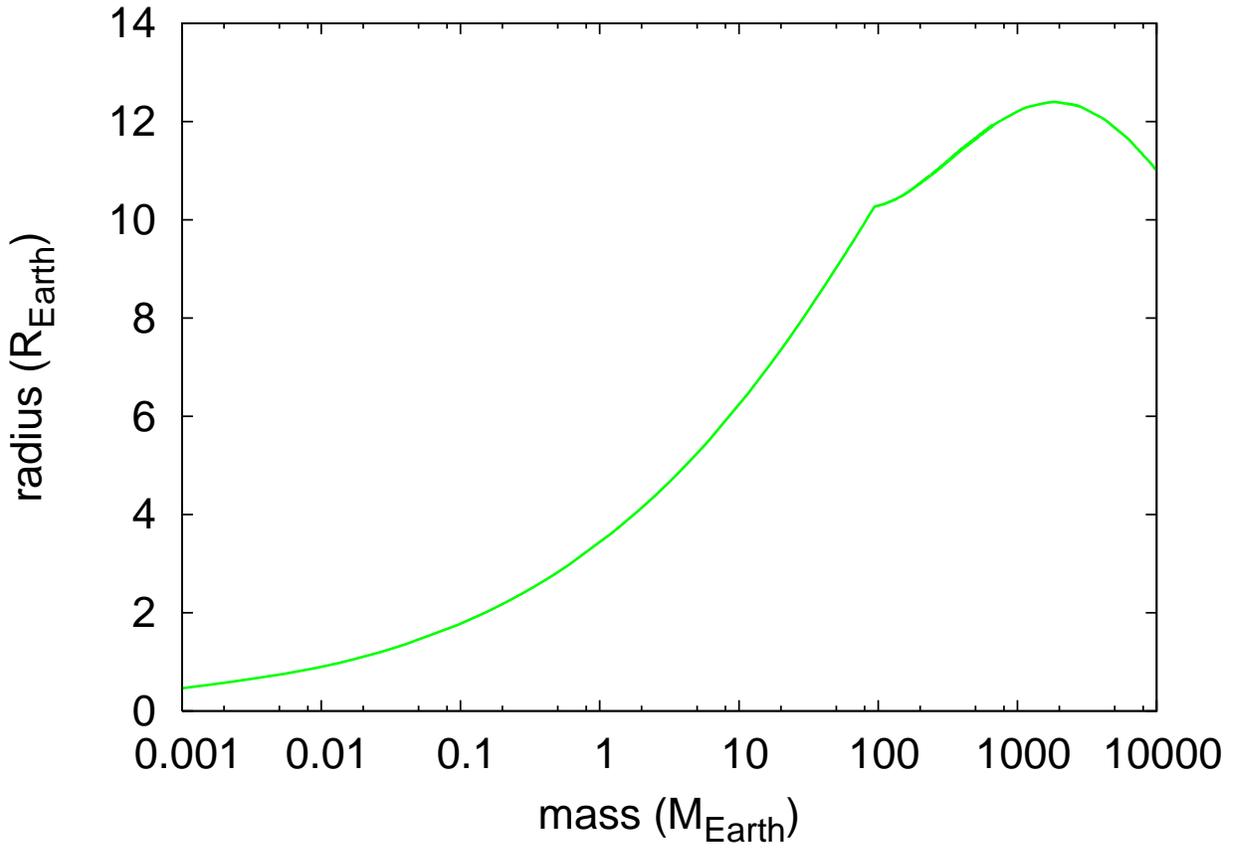}\end{center}
\caption{Mass-radius relation calculated for H.}
\label{fig:Hmr}
\end{figure}

\begin{figure}
\begin{center}\includegraphics[width=\textwidth]{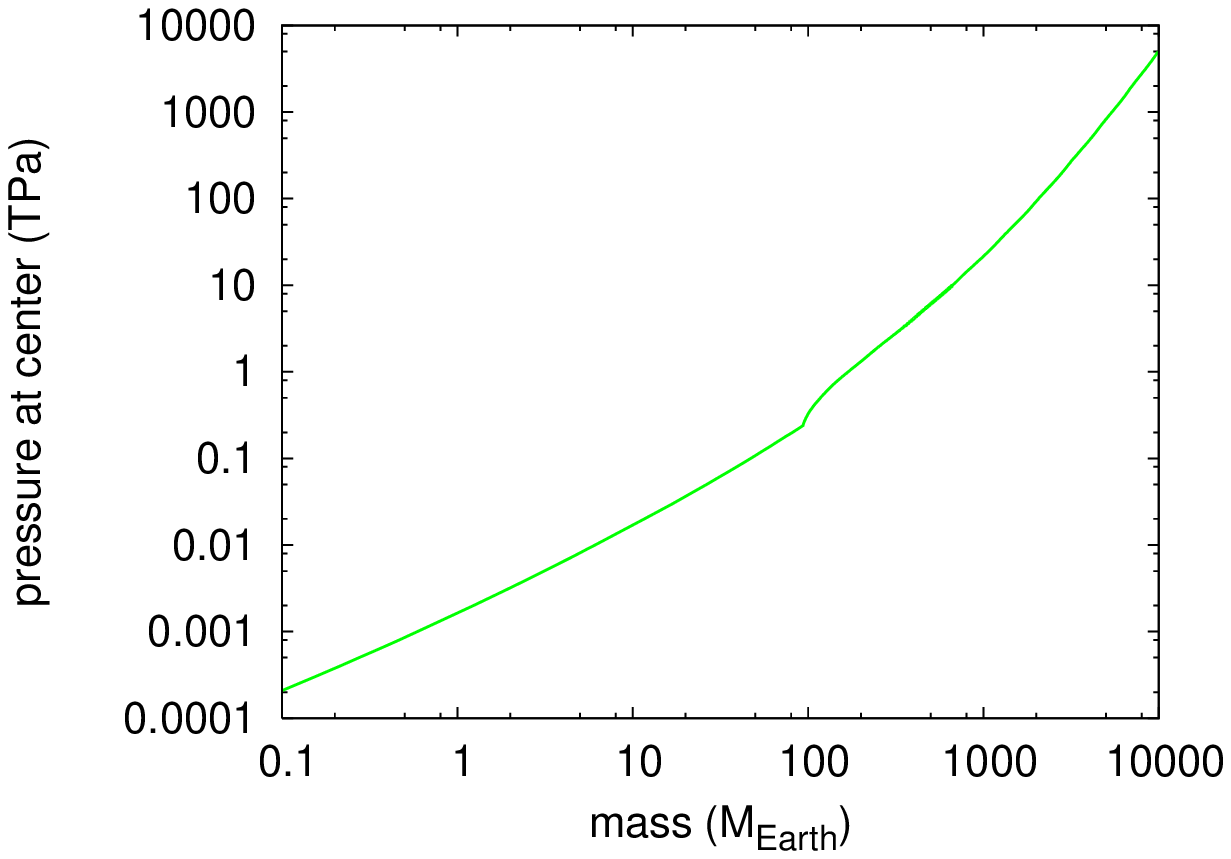}\end{center}
\caption{Predicted variation of core pressure with mass for H.}
\label{fig:Hmp}
\end{figure}

We did not find a H/He EOS of similar pedigree to the other EOS above,
so the mass-radius relation was calculated for pure H
(SESAME EOS 5251: \cite{AlbersJohnson1982}) as a limiting case of low density.
This EOS was isotopically scaled from SESAME EOS 5263 for D$_2$,
which is wide-ranging and includes dissociation
and ionization \citep{Kerley1972},
and has been found to give good agreement with static and shock
compression data.
As with the ices, the isentrope was chosen such that a well-defined
planetary surface was calculated, and is thus a lower bound on the radius
of planets with a thick atmosphere.
For the EOS used,
the isentrope passed through 10\,K at zero pressure,
and exhibited a kink at around 100\,\ME\ caused by metallization
at around 200\,GPa.
Interestingly, the resulting mass-radius relation is quite close to the 
TFD prediction \citep{Seager2007}.
Our calculation has a maximum in radius, 79000\,km
at a mass of $11\times 10^{16}$\,kg.
This result is slightly higher in mass and lower in radius than the
TFD prediction.
(Figs~\ref{fig:Hmr} and \ref{fig:Hmp}.)

\subsection{Fe/basalt}
\begin{figure}
\begin{center}\includegraphics[width=\textwidth]{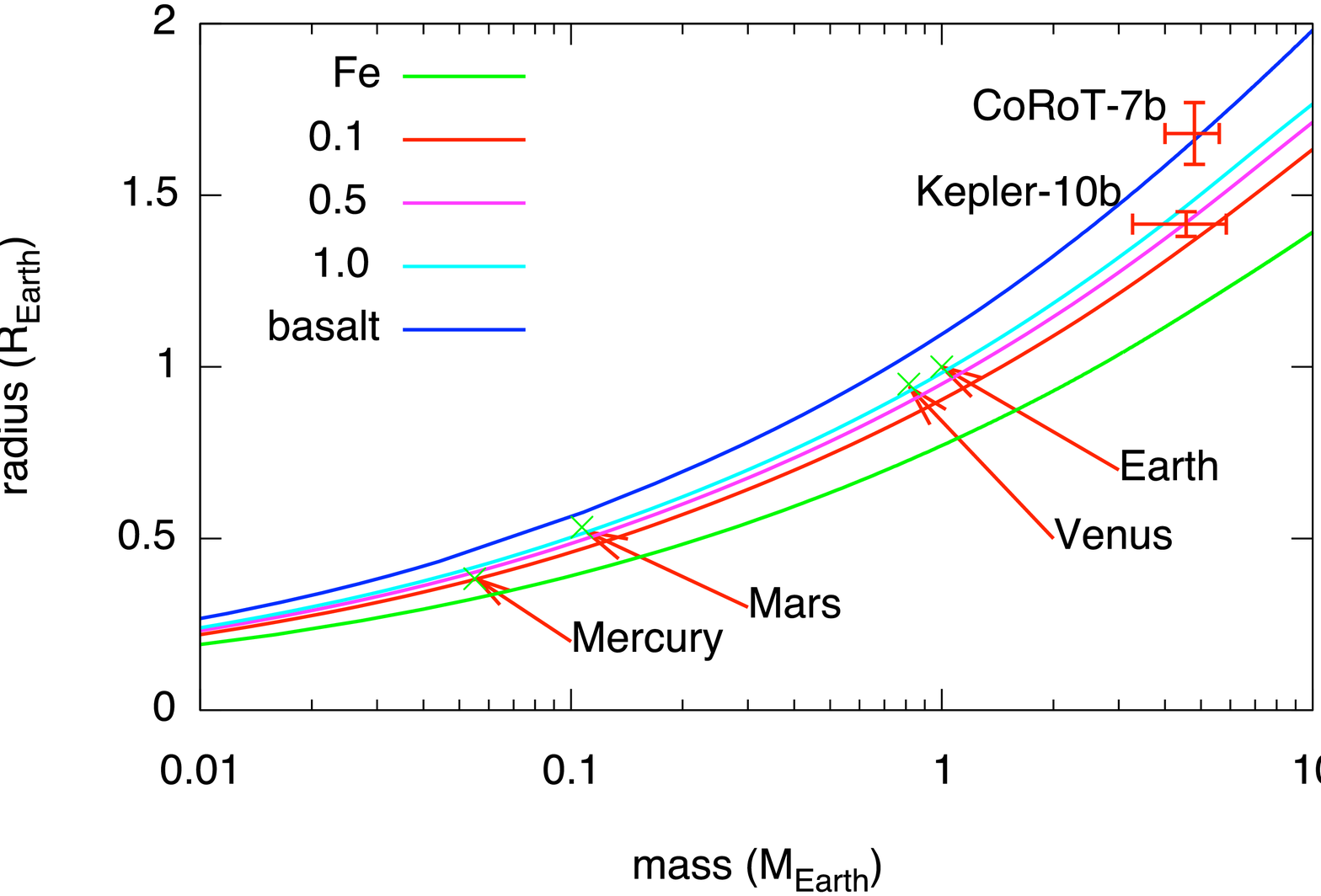}\end{center}
\caption{Mass-radius relations calculated for differentiated Fe/basalt planets,
   in terms of the ratio of basalt to Fe.}
\label{fig:Febasaltmr}
\end{figure}

Mass-radius relations were calculated for differentiated two-component planets
comprising an Fe core and a basalt (composition) mantle, with
fixed mass ratios, using the elliptic method of solution.
Fe and basalt were represented by SESAME EOS 2150 and 7530 respectively.
Because of its low density, the addition of an outer basalt layer gave a 
disproportionate change in the radius, to a greater extent than indicated in
previous work \citep{Seager2007}.
(Fig.~\ref{fig:Febasaltmr}.) 

\section{Application to exoplanets}
\begin{figure}
\begin{center}\includegraphics[width=\textwidth]{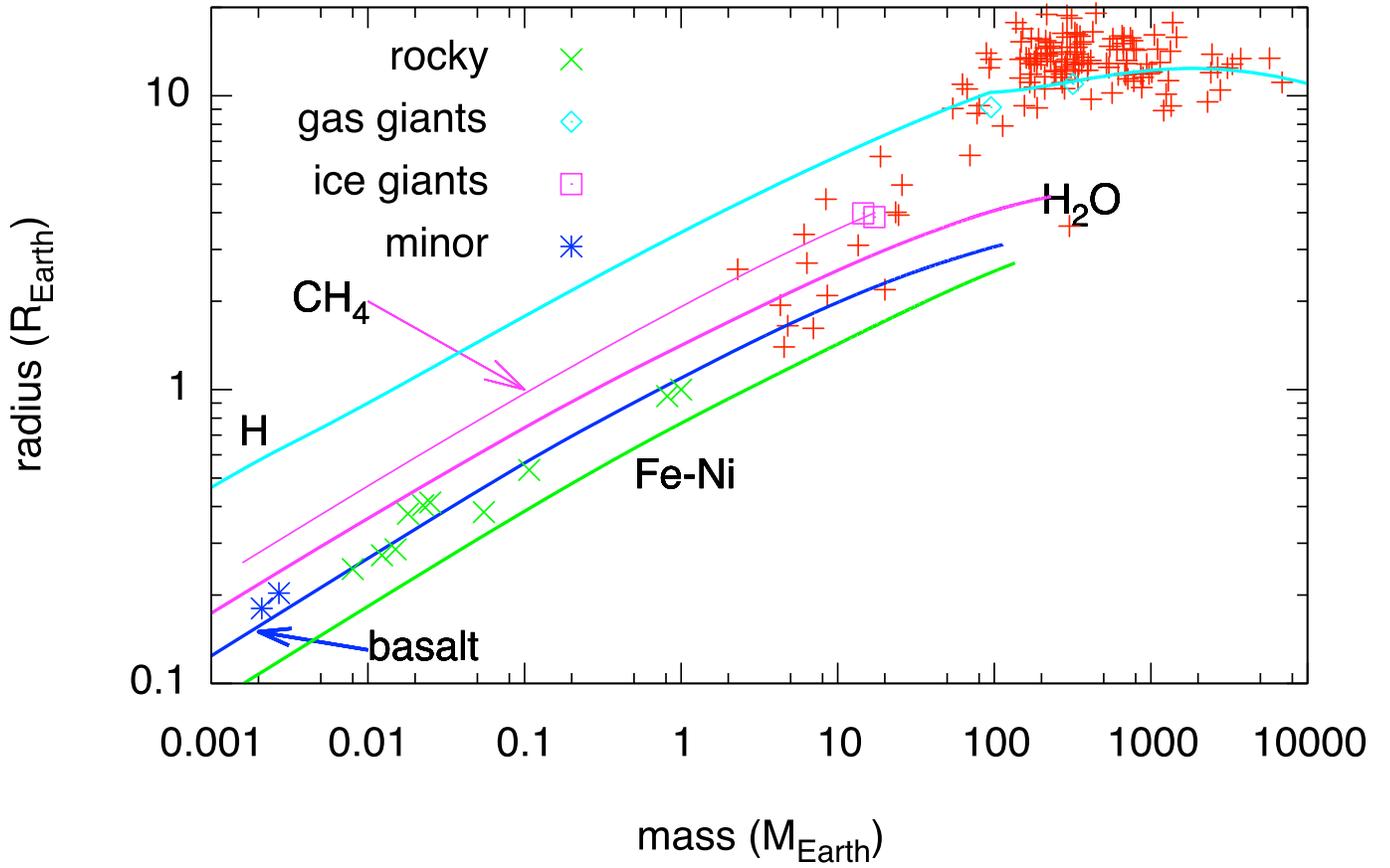}\end{center}
\caption{Mass-radius relations for different classes of material,
   compared with planets, moons, and exoplanets (red crosses).}
\label{fig:mrcmp2}
\end{figure}

\begin{figure}
\begin{center}\includegraphics[width=\textwidth]{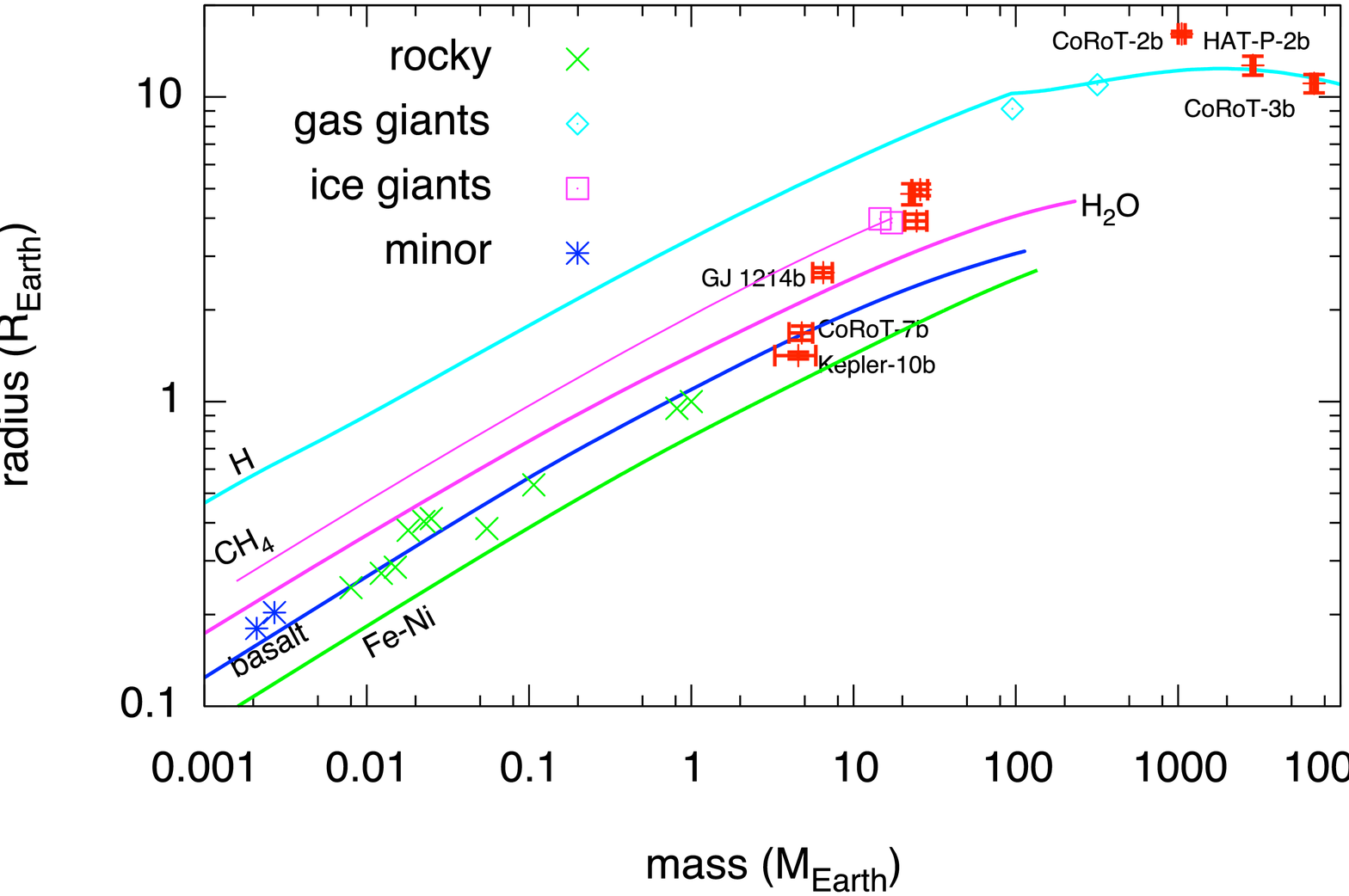}\end{center}
\caption{Mass-radius relations for different classes of material,
   compared with selected exoplanets.}
\label{fig:mrcmp}
\end{figure}

For comparative purposes, we chose to represent each type of
material by a single mass-radius curve corresponding to a single EOS:
Fe-Ni by the electronic structure cold curve for Fe$_3$Ni
(which is similar to the Fe SESAME 2150 curve but extends to higher pressure), 
rock by SESAME 5530 for basalt,
ice by SESAME 7154 for H$_2$O,
and H/He by SESAME 5251 for H.
Plotting the planets of the solar system, they
fall in the expected places with respect to the curves:
the rocky planets between Fe-Ni and rock with Mercury closest to Fe-Ni
and Mars closest to rock,
Jupiter and Saturn close to the H/He curve,
and Uranus and Neptune close to the CH$_4$ curve.
Minor planets Pluto and Eris are also shown,
lying on the icy side of the basalt curve.
We also compare exoplanets at extremes of size and mass:
`super-Earths' Kepler-10b \citep{Batalha2011},
CoRoT-7b \citep{Leger2009}, and GJ~1214b \citep{Charbonneau2009}
and the `super-Jupiters' HAT-P-2b \citep{Loeillet2008} 
and CoRoT-3b \citep{Deleuil2008},
whose core pressures
are representative of states it would be useful to explore in
future experiments on material properties,
and the anomalously large CoRoT-2b.
(Figs~\ref{fig:mrcmp2} and \ref{fig:mrcmp};
exoplanet parameters from \cite{Schneider2011}.)

Measurements of exoplanet Kepler-10b constrain
its radius to $4.56_{-1.29}^{+1.17}$\,$M_E$
and mass to $1.416_{-0.036}^{+0.033}$\,$R_E$.
The nominal value lies on our mass-radius relation for
Fe-basalt differentiated planets with 2/3 of the total mass in the core.
With this structure, the central pressure would be
$2.5_{-0.3}^{+0.4}$\,TPa.
Considering two-layer differentiated structures at the $1\sigma$ level,
the planet could have a core mass fraction between 25\%\ and 95\%,
with central pressure between about 1.5 and 3.7\,TPa.

Measurements of exoplanet CoRoT-7b
constrain its radius to $1.68\pm 0.09$\,\RE\ and mass to $4.8\pm 0.8$\,\ME
\citep{Leger2009},
though other researchers have deduced different,
and inconsistent, masses from the same Doppler shift data
\citep{Hatzes2010,Pont2011,Ferraz-Mello2011}.
Given the current discrepancy in mass deduced by different groups,
the validity of deduced compositions and central pressures is unclear,
but it is informative to assess the values and uncertainties derived
from the quoted uncertainties in mass and radius.
The nominal value by \cite{Leger2009} lies very close to our mass-radius
relation for basalt, for which the central pressure would be
$0.75_{-0.12}^{+0.14}$\,TPa.
Considering two-layer differentiated structures at the $1\sigma$ level,
the planet could be Fe/rock with a core of radius up to 4500\,km,
or rock/ice with a surface layer of (for example) H$_2$O up to 2500\,km thick.
The central pressure is calculated to increase rapidly with Fe,
giving $0.8_{-0.1}^{+1.2}$\,TPa.
For three-layer differentiated structures, the ice layer would be thicker
as the core size increased.
The uncertainty in deduced composition, core radius, and central pressure is 
nonlinear with respect to the uncertainty in mass and radius.
Measurements of Doppler shift in the stellar spectrum
and dimming in the stellar brightness are essentially independent,
so we calculate uncertainty contours in radius and central pressure
and smooth them to infer the $1\sigma$ uncertainty.
The planet is very unlikely to be Fe-only, in agreement
with the previous conclusions \citep{Leger2009}.
In fact, CoRoT-7b is likely to have a smaller core in proportion to its size 
than do the terrestrial planets.
The presence of ice is thought to be unlikely because of the proximity
of the planet to its star, CoRoT-7, resulting in high surface temperatures.
The nominal mass-radius measurement is consistent with a composition of
rock only;
given the uncertainties in the mass-radius measurement and relations,
a metallic core cannot be ruled out.

Recent measurements of exoplanet GJ~1214b give 
a radius of $2.65\pm 0.11$\,\RE\ and mass of 
$6.55\pm 0.9$\,\ME\ \citep{Charbonneau2009}
or $6.45\pm 0.9$\,\ME\ \citep{Carter2011},
lying between the mass-radius curves for H$_2$O and CH$_4$.
The closest analogue in the solar system is the icy giants;
the mass and radius of GJ~1214b are consistent with a higher proportion of 
H$_2$O, or with a proportionately larger amount of rock or Fe,
than Neptune or Uranus.
The observations do not of themselves suggest an `ocean planet'
as the planet's density seems implausibly low.
Furthermore, as has been pointed out previously \citep{Adams2008}, 
a mean density consistent with a `water world' could arise
from a rocky interior with an H/He atmosphere.

Recent parameters for the super-massive exoplanet HAT-P-2b are 
a mass of $9.09\pm 0.24$\,\MJ\ 
and a radius of $1.157_{-0.092}^{+0.073}$\,\RJ\ \citep{Pal2009}.
These parameters are consistent with a massive hydrogen-rich planet,
as concluded previously.
Proportionately, the radius is smaller with respect to our predicted
mass-radius curve for hydrogen than are Jupiter or Saturn, suggesting a
larger proportion of rock or Fe.
A pure hydrogen planet of the nominal mass would have a radius
of 1.12\,MJ and a central pressure of 214\,TPa.
The actual proportion of rock that would be consistent with the
observed mass and radius is difficult to estimate using the available EOS,
because they roll over at lower masses;
future studies will include theoretical prediction and experimental
investigation of the EOS under relevant conditions.

The super-massive exoplanet CoRoT-3b has a mass of $21.7\pm 1$\,\MJ\
and $1.01\pm 0.07$\,\RJ\ \citep{Deleuil2008},
lying close to our mass-radius curve for H.

Our mass-radius relation for H confirms the anomalously large radius
of CoRoT-2b.
As proposed recently \citep{Leconte2009}, a plausible explanation is that
this planet is young and hot, from accretion, tidal heating,
or internal processes. 
The radius would be larger for a warmer isentrope,
but is relatively insensitive to temperature.
Another potential source of uncertainty, particularly for warm
bodies with a significant proportion of volatiles,
is the sensitivity to definition of the outer surface of the body.
If the low-density gas in the outer region of the atmosphere
has a higher opacity than expected, the planet would appear larger.

\begin{figure}
\begin{center}\includegraphics[width=\textwidth]{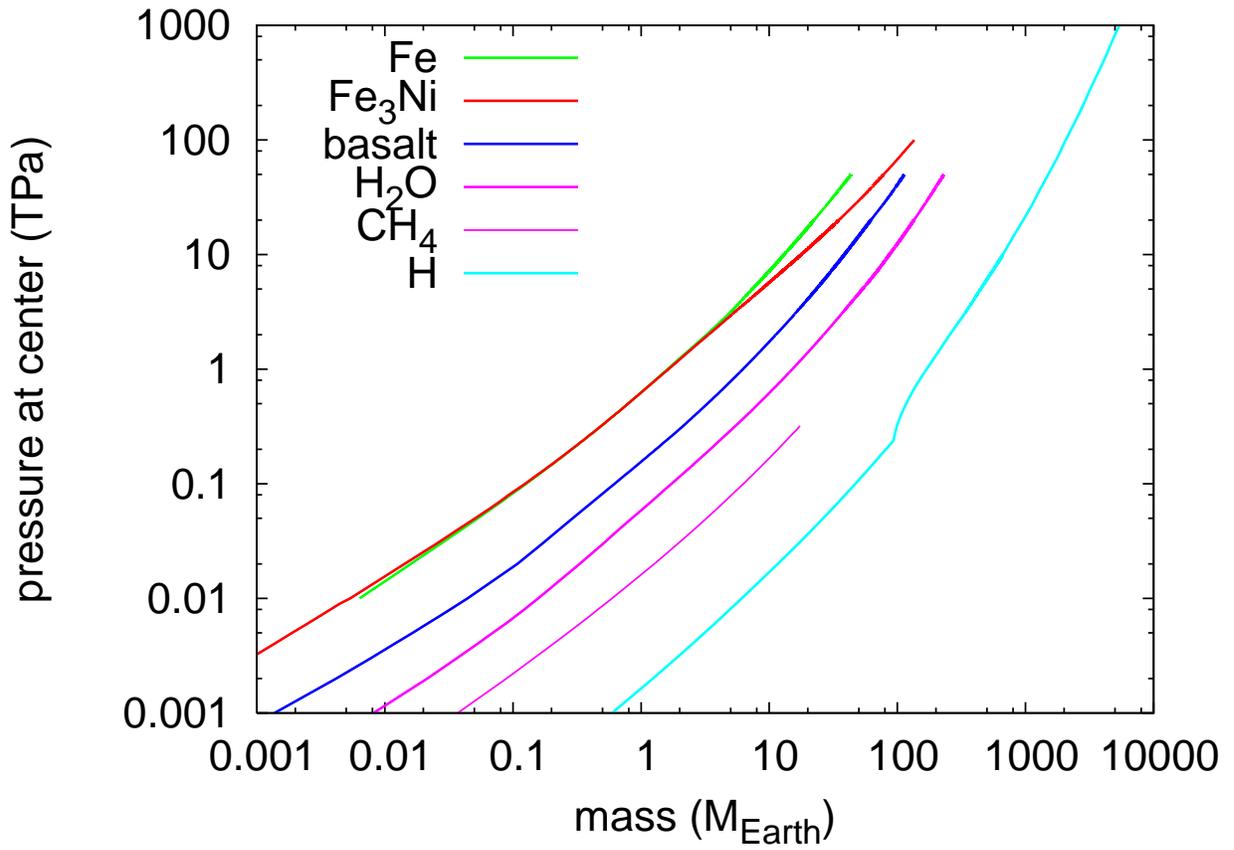}\end{center}
\caption{Mass-pressure relations for different classes of material.}
\label{fig:mpcmp}
\end{figure}

Using the location of a given planet with respect to the mass-radius 
relation for different compositions, it is possible to estimate the
pressure in the center of the core from the corresponding mass-pressure relation
(Fig.~\ref{fig:mpcmp}).
Thus the central pressure of Kepler-10b is likely to be between 1.5 and 2.7\,TPa,
and that of CoRoT-7b is likely to be between 0.7 and 2\,TPa,
which are accessible using planar ramp-loading experiments at the OMEGA
and NIF facilities.
The core pressure in GJ~1214b is approximately 1\,TPa: the same mass of H$_2$O 
gives a pressure of around 1.5\,TPa, and that mass of CH$_4$, 0.5\,TPa.
It would be most useful to improve EOS of ice compositions
into this regime, by QMD simulations and dynamic loading experiments;
again, planar ramp experiments at NIF should be able to access relevant states.
The core pressure in HAT-P-7b is approximately 20\,TPa,
which may be accessible with subsequent NIF experiments using
convergent compression.
The core of CoRoT-3b is at the highest pressure known for 
non-stellar matter: 1900\,TPa according to the EOS used.
Such pressures are achieved in inertial confinement fusion experiments,
but will require very significant experimental developments before
EOS experiments can be performed.

\section{Conclusions}
Mass-radius relations were derived using EOS widely used for, 
and thus validated against, dynamic compression experiments.
The results were similar to relations used previously to infer planetary 
structures, but were not identical.

Mass-radius relations for Fe were inferred using EOS calculated 
to higher pressures than reported previously, using density functional theory.
The difference between the cold curve and STP isentrope had a negligible
effect on the mass-radius relation in the exoplanet regime.
The effect of Ni in Fe was significant for the EOS, 
but negligible for the mass-radius relation of exoplanets.
In a differentiated planet,
the radius was more sensitive to small proportions of rock than has been
concluded previously.

These results support the conclusion that CoRoT-7b,
with measurements analyzed by \cite{Leger2009}, is `Earth-like' 
in that it is likely to have a rocky layer over an Fe-Ni core,
though the present uncertainty in the exoplanet's mass is too large
to constrain the mass of Fe significantly: there may be very little.
GJ~1214b lies between the curves for H$_2$O and CH$_4$, which are
compositions where improvements to the EOS are particularly desirable.
The composition cannot be constrained meaningfully from present data,
and suggestions that it might be an `ocean planet' are speculative.
The present results also support the conclusion that HAT-P-2b is a gas giant,
though the density is proportionately larger than the gas giants in 
the solar system, indicating a higher proportion of matter of
higher atomic number.
The mass-radius relations imply
core pressures of $0.75\pm 0.15$\,TPa for CoRoT-7b,
a range range 0.5-1.5\,TPa for GJ~1214b, 
over 210\,TPa in HAT-P-2b, and around 1900\,TPa in CoRoT-3b.

Based on the constituent physics and approximations,
we conclude that current theoretical techniques for electronic structure, 
with care in the use of pseudopotentials,
should be as suitable for predicting EOS in the exoplanet regime as they are at
low pressures.
Shock measurements are relevant in terms of the compressions that can be 
explored, though interpretation is involved to account for the elevated
temperatures.
Ramp loading techniques are very appropriate, and planned NIF experiments
should provide data on Fe in the region where DFT mass-radius relations
deviate from SESAME EOS.
In both experiment and theory, 
the limiting cases of Fe (or Fe-Ni) and H are the subject of active research.
Rocks and ices are more complicated in terms of composition and structure,
and there is clearly a need for future research into the properties of
these materials at compressions relevant for planets and exoplanets.

\section*{Acknowledgments}
The authors would like to acknowledge the invaluable contributions of
G.I.~Kerley, J.D.~Johnson at Los Alamos National Laboratory, 
W.M.~Howard at Lawrence Livermore National Laboratory,
and K.~Rice at the Royal Observatory, Edinburgh.
This work was performed
under the auspices of
the U.S. Department of Energy under contract
DE-AC52-07NA27344.


\begin{thebibliography}{}
\bibitem[Ackland(2002)]{Ackland2002}{Ackland, G.J. 2002,
   J. Phys.: Condens. Matter, 14, 2975}
\bibitem[Adams et al(2008)]{Adams2008}{Adams, E.R., Seager, S., \& Elkins-Tanton, L.
   2008,
   ApJ, 673, 1160}
\bibitem[Ahrens \& Gregson(1964)]{Ahrens1964}{Ahrens, T.J., \& Gregson, J.V.G.
   1964,
   J. Geophys. Res., 69, 4839-4874}
\bibitem[Ahrens(1966)]{Ahrens1966}{Ahrens, J.J. 1966,
   J. Appl. Phys., 37, 2532-2541}
\bibitem[Albers \& Johnson(1982)]{AlbersJohnson1982}{Albers, R. \& Johnson, J.D. 1982,
   documentation for SESAME EOS 5251 for H, in \cite{SESAME}}
\bibitem[Alf\`e et al(2000)]{Alfe2000}{Alf\`e, D., Gillan, M.J., \& Price, G.D. 2000,
   Nature, 405, 172}
\bibitem[Alf\`e et al(2000a)]{Alfe2000a}{Alf\`e, D., Kresse, G. \& Gillan, M.J. 2000,
   Phys. Rev. B, 61, 132}
\bibitem[Alf\`e et al(2001)]{Alfe2001}{Alf\`e, D., Price, G.D., \& Gillan, M.J. 2001,
   Phys. Rev. B, 64, 045123}
\bibitem[Barnes \& Rood(1973)]{BarnesRood1973}{Barnes, J. \& Rood, J. 1973,
   (Los Alamos National Laboratory),
   documentation for SESAME EOS 2140 for Fe, in \cite{SESAME}}
\bibitem[Barnes et al(1974)]{Barnes1974}{Barnes, J.F. et al 1974,
   J.~Appl. Phys., 45, 727}
\bibitem[Barnes \& Lyon(1988a)]{BarnesLyon1988_MgO}{Barnes, J.F. \& Lyon, S.P. 1988,
   (Los Alamos National Laboratory),
   documentation for SESAME EOS 7460 for MgO, in \cite{SESAME92}}
\bibitem[Barnes \& Lyon(1988b)]{BarnesLyon1988_basalt}{Barnes, J.F. \& Lyon, S.P. 1988,
   {\it SESAME Equation of State Number 7530, Basalt},
   Los Alamos National Laboratory report LA-11253-MS}
\bibitem[Bastea et al(2007)]{CHEETAH}{Bastea, S., Fried, L.E., Howard, W.M., Souers, P.C.,
   \& P.A.~Vitello, P.A. 2007,
   CHEETAH computer program (current version is 5.0,
   Lawrence Livermore National Laboratory release number
      UCRL-CODE-234077)}
\bibitem[Belonoshko et al(2008)]{Belonoshko2008}{%
   Belonoshko, A.B., Dorogokupets, P.I., Johansson, B., Saxena, S.K., \& Koi, L. 2008,
   Phys. Rev. B, 78, 104107}
\bibitem[Batalha et al(2011)]{Batalha2011}{Batalha, N.M. et al 2011,
  ApJ, 729, 27}
\bibitem[Bradley et al(2009)]{Bradley2009}{Bradley, D.K. et al 2009,
   Phys. Rev. Lett., 102, 075503}
\bibitem[Bushman et al(1993)]{Bushman1993}{Bushman, A.V., Kanel', G.I., Ni, A.L.,
   \& Fortov, V.E. 1993
   Intense Dynamic Loading of Condensed Matter
   (Taylor \& Francis, London)}
\bibitem[Caracas \& Cohen(2008)]{Caracas2008}{Caracas, R. \& Cohen, R.E. 2008,
   PEPI, 168, 3-4, 147 and references therein}
\bibitem[Carter et al(2011)]{Carter2011}{Carter, J.A., Winn, J.N., Holman, M.J.,
   Fabricky, D., Berta, Z.K., Burke, C.J., and Nutzman, P. 2011,
   ApJ, 730, 82}
\bibitem[Charbonneau et al(2000)]{Charbonneau2000}{Charbonneau, D. 
   Brown, T.M., Latham, D.W., \& Mayor, M. 2000,
   ApJ, 529, L45-48}
\bibitem[Charbonneau et al(2002)]{Charbonneau2002}{Charbonneau, D., Brown, T.M., Noyes, R.W.,
   \& Gilliland, R.L. 2002,
   ApJ, 568, 1, 377}
\bibitem[Charbonneau et al(2009)]{Charbonneau2009}{Charbonneau, D. et al 2009,
   Nature, 462, 891}
\bibitem[Chau et al(2011)]{Chau2011_ice}{Chau, R., Hamel, S., \& Nellis, W.J.
   2011, Nature Comm., 2, 203}
\bibitem[Clerouin(2002)]{Clerouin2002}{Clerouin, J. 2002, 
   J. Phys.: Condens. Matter, 14, 9089}
\bibitem[Delaney et al(2006)]{Delaney2006}{Delaney, K., Pierleoni, C. \& Ceperley, D.M. 2006,
   Phys. Rev. Lett., 97, 235702-1-4}
\bibitem[Deleuil et al(2008)]{Deleuil2008}{Deleuil, M. et al 2008,
   A\&A, 491, 889}
\bibitem[Drummond \& Ackland(2002)]{Drummond2002}{Drummond, N.D., \& Ackland, G.J. 2002,
   Phys. Rev. B, 65, 184104}
\bibitem[Edwards et al(2004)]{Edwards2004}{Edwards, J. et al 2004,
   Phys. Rev. Lett., 92, 075002}
\bibitem[Ferraz-Mello et al(2011)]{Ferraz-Mello2011}{Ferraz-Mello, S., et al 2011,
   A\&A, 531, A161}
\bibitem[Fortney et al(2007)]{Fortney2007}{Fortney, J.J., Marley, M.S.,
   \& Barnes, J.W. 2007,
   ApJ, 668, 1267}
\bibitem[Grasset et al(2009)]{Grasset2009}{Grasset, D., Schneider, J., \& Sotin, C. 2009,
   ApJ, 693, 722}
\bibitem[Hamel et al(2011)]{Hamel2011}{Hamel, S., Morales, M., \& Schwegler, E. 2011,
   Phys. Rev. B (in press)}
\bibitem[Henry et al(2000)]{Henry2000}{Henry, G.W. et al. 2000,
   ApJL, 529, 1, L41}
\bibitem[Hohenberg \& Kohn(1964)]{Hohenberg1964}{Hohenberg, P. \& Kohn, W. 1964,
   Phys. Rev. B, 136, 3B, 864}
\bibitem[Holian(1984)]{SESAME}{Holian, K.S. (Ed.) 1984, Los Alamos National Laboratory
   report LA-10160-MS Vol 1c}
\bibitem[Holst et al(2011)]{Holst2011}{Holst, B., French, M, and Redmer, R. 2011,
   Phys. Rev. B, 83, 235120}
\bibitem[Haar \& Gallagher(1978)]{Johnson1982_5520}{Haar, L. \& Gallagher, J.S. 1978,
   J.~Phys. Chem. Ref. Data, 7, 3, 635}
\bibitem[Hatzes et al(2010)]{Hatzes2010}{Hatzes, A. et al 2010,
   A\&A, 520, A93}
\bibitem[Hensel(1999)]{Hensel1999}{Hensel, F. 1999,
   J. Non-Crystalline Solids, 150-152, 135-138}
\bibitem[Hicks et al(2005)]{Hicks2005}{Hicks, D.G., Boehly, T.R., Celliers, P.M.,
   Eggert, J.H., Vianello, E., Meyerhofer, D.D., \& Collins, G.W. 2005,
   Phys. Plasmas, 12, 082702}
\bibitem[Johnson(1984)]{Johnson1984_5502}{Johnson, J.D. 1984,
   documentation for SESAME EOS 5502 for CH$_4$,
   in \cite{SESAME}}
\bibitem[Johnson \& Lyon(1984)]{JohnsonLyon1984}{Johnson, J.D. \& Lyon, S.P. 1984,
   documentation for SESAME EOS 7383 for SiO$_2$, in \cite{SESAME92}}
\bibitem[Johnson \& Lyon(1990)]{JohnsonLyon1990}{Johnson, J.D. \& Lyon, S.P. 1990,
   documentation for SESAME EOS 7154 for H$_2$O, in \cite{SESAME92}}
\bibitem[Kerley(1972)]{Kerley1972}{Kerley, G.I. 1972,
   Los Alamos National Laboratory report LA-4776}
\bibitem[Kerley(1980)]{Kerley1980}{Kerley, G.I. 1980,
   J.~Appl. Phys., 51, 10, 5369}
\bibitem[Kerley(1993)]{Kerley1993}{Kerley, G.I. 1993,
   Sandia National Laboratories report SAND93-0227}
\bibitem[Kittel \& Kroemer(1980)]{Kittel1980}{Kittel, C. \& Kroemer, H. 1980,
   Thermal Physics (W.H.~Freeman, New York).}
\bibitem[Knudson \& Desjarlais(2009)]{Knudson2009}{Knudson, M.D., \&
   Desjarlais, M.P. 2009,
   Phys. Rev. Lett., 103, 225501}
\bibitem[Kohn \& Sham(1965)]{Kohn1965}{Kohn, W. \& Sham, L.J. 1965,
   Phys. Rev., 140, 4A}
\bibitem[Leconte et al(2009)]{Leconte2009}{Leconte, J., Baraffe, I., Chabrier, G., Barman, T.,
   \& Levrard, B. 2009,
   A\&A, 506, 385}
\bibitem[L\'eger et al(2009)]{Leger2009}{L\'eger, A. et al 2009,
   A\&A manuscript 11933,
   preprint {\tt arXiv:0908.0241v3}}
\bibitem[Liberman(1979)]{Liberman1979}{Liberman, D.A. 1979,
   Phys. Rev. B, 20, 12, 4981}
\bibitem[Lindl(1998)]{ICF}{Lindl, J.D. 1998,
   Inertial Confinement Fusion
   (Springer-Verlag, New York)}
\bibitem[Loeillet et al(2008)]{Loeillet2008}{Loeillet, B. et al 2008,
   A\&A, 481, 529}
\bibitem[Lorenzen et al(2010)]{Lorenzen2010}{Lorenzen, W, Holst, B,
   \& Redmer, R. 2010,
   Phys. Rev. B, 82, 195107}
\bibitem[Luo et al(2004)]{Luo_MgO_2004}{Luo, S.-N., Swift, D.C., Mulford, R.N.,
   Drummond, N.D., \& Ackland, G.J. 2004,
   J. Phys.: Cond. Matt., 16, 30, 5435}
\bibitem[Lyon \& Johnson(1992)]{SESAME92}{Lyon, S.P. \& Johnson, J.D. 1992,
   Los Alamos National Laboratory report LA-UR-92-3407}
\bibitem[Mayor \& Queloz(1995)]{Mayor1995}{Mayor, M. \& Queloz, D. 1995,
   Nature, 378, 355-359}
\bibitem[McQueen et al(1970)]{shock}{%
   McQueen, R.G., Marsh, S.P., Taylor, T.W., Fritz, J.N., \& Carter, W.J. 1970,
   in High Velocity Impact Phenomena,
   ed. R.~Kinslow,
   (Academic Press, New York)}
\bibitem[Michielsen \& De~Raedt(1996)]{Michielsen1996}{Michielsen, K. \& De~Raedt, H. 1996,
   Europhys. Lett., 34, 6, 435}
\bibitem[Militzer(2009)]{Militzer2009}{Militzer, B. 2009,
 Phys. Rev. B, 79, 155105-1-18}
\bibitem[Morales et al(2009)]{Morales2009}{Morales, M.A., Schwegler, E., Ceperley, D.M., 
   Pierleoni, C., Hamel, S., \& Caspersen, K. 2009,
   PNAS, 106, 1324 and references therein}
\bibitem[Morales et al(2010)]{Morales2010}{Morales, M.A., Pierleoni, C.,
   Schwegler, E., \& Ceperley, D.M. 2010,
hydrogen from ab initio simulations
   PNAS, 107, 12799-12803}
\bibitem[Mulford \& Swift(2006)]{MulfordSwift2006}{Mulford, R.N. \& Swift, D.C. 2006,
   Proc. Am. Phys. Soc. Topical Conf. on Shock Compression of Cond. Matt.,
   Conf. Proc. 845 pp.461-464 (Am. Inst. Phys.)}
\bibitem[P\'al et al(2009)]{Pal2009}{P\'al, A. et al 2009,
   MNRAS, 401, 4, 2665-2674}
\bibitem[Payne et al(1992)]{Payne1992}{Payne, M.C., Teter, M.P., Allan, D.C., Arias, T.A. \&
   Joannopoulos, J.D. 1992,
   Rev. Mod. Phys., 64, 4}
\bibitem[Perdew(1992)]{Perdew1992}{Perdew, J. 1992, Phys. Rev., B46, 6671}
\bibitem[Perdew(1994)]{Perdew1994}{Perdew, J. 1994, Phys. Rev., B50, 4954}
\bibitem[Pont et al(2011)]{Pont2011}{Pont, F., Aigrain, S., \& Zucker, S. 2011,
   MNRAS, 411, 1953}
\bibitem[Press et al(1989)]{Press1989}{Press, W.H., Flannery, B.P., Teukolsky, S.A. \&
   Vetterling, W.T. 1989,
   Numerical Recipes (Cambridge University Press, London)}
\bibitem[Ragan(1984)]{Ragan1984}{Ragan III, C.E. 1984
   Phys. Rev. A, 25, 6, 3360}
\bibitem[Recoules et al(2009)]{Recoules2009}{
   Recoules, V., Lambert, F., Decoster, A., Canaud, B., \& Clerouin, J. 2009,
   Phys. Rev. Lett., 102, 075002-1-4}
\bibitem[Reisman et al(2001)]{Reisman2001}{Reisman, D.B. et al 2001,
   J.~Appl.~Phys., 89, 1625}
\bibitem[Salpeter \& Zapolsky(1967)]{Salpeter1967}{Salpeter, E.E., \&
   Zapolsky, H.S. 1967,
   Phys. Rev., 158, 876}
\bibitem[Saumon \& Chabrier(1992)]{Saumon1992}{Saumon, D. \& Chabrier, G. 1992
   Phys. Rev. A, 46, 2084}
\bibitem[Schneider(2011)]{Schneider2011}{Schneider, J. 2011,
   The Extrasolar Planets Encyclopaedia (version 2.06),
   {\tt http://exoplanet.eu} and references therein}
\bibitem[Seager et al(2007)]{Seager2007}{Seager, S., Kuchner, M., Hier-Majumder, C.A., \& Militzer, B. 2007
   ApJ, 669, 1279}
\bibitem[Sola et al(2009)]{Sola2009}{Sola, E., Brodholt, J.P., \& Alf\`e, D. 2009,
   Phys. Rev. B, 79, 024107}
\bibitem[Stacey \& Davis(2004)]{Stacey2004}{Stacey, F.D., \& Davis, P.M. 2004,
   PEPI, 142, 137-184}
\bibitem[Stixrude \& Cohen(1995)]{Stixrude1995}{Stixrude, L. \& Cohen, R.E. 1995,
   Science, 267, 1972}
\bibitem[Stixrude et al(2009)]{Stixrude2009}{Stixrude, L., de Koker, N.,
   Mookherjee, M., \& Karki, B.B. 2009,
   EPSL, 278, 226-232}
\bibitem[Stixrude \& Lithgow-Bertelloni(2010)]{Stixrude2010}{Stixrude, L. \&
   Lithgow-Bertelloni, C. 2010,
   Rev. Mineralogy \& Geochem., 71, 465}
\bibitem[Stixrude \& Lithgow-Bertelloni(2011)]{Stixrude2011}{Stixrude, L. \&
   Lithgow-Bertelloni, C. 2011,
   Geophys. J. Int., 184, 1180}
\bibitem[Swift et al(2001)]{Swift_SiEOS_2001}{Swift, D.C., Ackland, G.J., Hauer, A., \& Kyrala, G.A. 2001,
   Phys. Rev. B, 64, 214107}
\bibitem[Swift \& Johnson(2005)]{Swift_lice_2005}{Swift, D.C. \& Johnson, R.P. 2005,
   Phys. Rev. E, 71, 066401}
\bibitem[Swift et al(2005)]{Swift_NiTi_2005}{Swift, D.C. et al 2005,
   J. Appl. Phys., 98, 093512}
\bibitem[Swift et al(2007)]{Swift_NiAlEOS_2007}{Swift, D.C., Paisley, D.L., McClellan, K.J.,
   \& Ackland, G.J. 2007,
   Phys. Rev. B, 76, 134111}
\bibitem[Swift(2008)]{Swift_genscalar_2008}{Swift, D.C. 2008,
   J. Appl. Phys., 104, 7, 073536}
\bibitem[Swift(2009)]{Swift_planetstruc_2009}{Swift, D.C. 2009,
   {\it Sensitivity of rocky planet structures to the equation of state},
   preprint {\tt arXiv:0908.3294}}
\bibitem[Swift(2010)]{Swift_Feeos_2010}{Swift, D.C. 2010,
   {\it Equations of state for Fe and Fe-Ni to massive exoplanet core conditions},
   in preparation}
\bibitem[Swift(2011)]{Swift2011_NH3EOS}{Swift, D.C. 2011,
   {\it Gr\"uneisen equation of state for ammonia},
   in preparation.}

\bibitem[Troullier \& Martin(1991)]{Troullier1991}{Troullier, N., \& Martins, J.L., 1991, Phys. Rev. B, 43, 1993}
\bibitem[Valencia et al(2010)]{Valencia2010}{Valencia, D., Ikoma, M., Guillot, T.,
   \& Nettelmann, N. 2010,
   A\&A, 516, A20}

\bibitem[van Thiel(1966)]{vanThiel1966}{van Thiel, M. 1966,
   {\it Compendium of Shock Wave Data},
   Lawrence Radiation Laboratory report UCRL-50108}
\bibitem[Wasserman et al(1996)]{Wasserman1996}{%
   Wasserman, E., Stixrude, L., \& Cohen, R.E. 1996,
   Phys. Rev. B, 53, 8296}
\bibitem[Wentzcovitch et al(2010)]{Wentzcovitch2010}{Wentzcovitch, R.M.,
   Wu, Z., \& Carrier, P. 2010,
   Rev. Mineralogy \& Geochem., 71, 99}
\bibitem[Wolszan \& Frail(1992)]{Wolszcan1992}{Wolszczan, A. \& D.A.~Frail 1992,
   Nature, 355, 145}
\end{thebibliography}
\end{document}